\documentclass[preprint, aps]{revtex4}
\usepackage{amssymb, amsmath}
\usepackage{indentfirst}
\usepackage[mathscr]{eucal}
\usepackage{graphicx}

\begin{document}

\title{Heat conductivity in the $\beta$-FPU lattice. \\
Solitons and breathers as energy carriers}

\author{T.Yu.~Astakhova}

\author{V.N.~Likhachev}

\author{G.A.~Vinogradov${}^*$}

\address{Emanuel Institute of Biochemical Physics RAS, \\
 ul.~Kosygina~4, Moscow~119334, Russian Federation}

\begin{abstract}

This paper consists of two parts. The first part proposes a new
methodological framework within which the heat conductivity in 1D
lattices can be studied. The total process of heat conductivity is
decomposed into two contributions where the first one is the
equilibrium process at equal temperatures $T$ of both lattice ends
and the second -- non-equilibrium process with the temperature
$\Delta T$ of one end and zero temperature of the other. This
approach allows to isolate and analyze the heat transfer in
explicit form. The heat conductivity in the limit  $\Delta T \to
0$ is reduced to the heat conductivity of harmonic lattice with
stochastic rigidities determined by the equilibrium process at
temperature $T$. A threshold temperature $T_{\rm thr}$ is found
which separates two regimes: small perturbations exponentially
decay at $T < T_{\rm thr}$ and tend to constant value at $T >
T_{\rm thr}$. The threshold temperature scales $T_{\rm thr}(N)
\sim N^{-3}$ with the lattice size $N$ and $T_{\rm thr} \to 0$ in
the thermodynamic limit. Some unusual properties of heat
conductivity can be exhibited on nanoscales at low temperatures.
The thermodynamics of the $\beta$-FPU lattice can be adequately
approximated by the harmonic lattice with temperature renormalized
coefficients of rigidity. The second part testifies in the favor
of the soliton and breather contribution to the heat conductivity
in contrast to conclusions made in [N. Li, B. Li, S. Flach, PRL
105 (2010) 054102]. In the long-wavelength continuum limit the
discrete $\beta$-FPU lattice is reduced to the modified Korteweg
-- de~Vries equation. This equation has soliton and breather
solutions. Numerical simulations demonstrate their high stability.
New method for the visualization of moving solitons and breathers
is suggested. An accurate expression for the dependence of the
sound velocity on temperature is also obtained. Our results
support the conjecture on the solitons and breathers contribution
to the heat conductivity. The fraction of total heat flux
transferred by solitons and breathers merits additional analysis.

\vspace{0.1 cm}


\noindent ${}^*$The corresponding author:
\texttt{gvin@deom.chph.ras.ru}.

\end{abstract}

\maketitle

%%%%%%%%%%%%%%%%%%%%%%%%%%%%%%%%%%%%%%%%%%%%%%%%%%

\section{Introduction}%

The problem of heat conductivity in low dimensional systems
attracts much attention in last decades (see review \cite{Lep03})
and is motivated by the discovery of quasi-one-dimensional
(nanotubes, nanowires, etc.) and two-dimensional (graphen,
graphan, etc.) systems.

The modern theory of heat conductivity was initiated by the
celebrated preprint of E.~Fermi, J.~Pasta, and S.~Ulam
\cite{Fer55}, though the primary aim was ``{\it of establishing,
experimentally, the rate of approaching to the equipartition of
energy among the various degrees of freedom}''. Subsequent
investigations demonstrated wide area of consequences in many
physical and mathematical phenomena (see reviews in special issues
of journals CHAOS \cite{Cha05} and Lecture Notes in Physics
\cite{Lec08} devoted to the 50th anniversary of the FPU preprint).

The dynamical properties of nonlinear systems in microcanonical
ensemble (total energy $E =$ const) were thoroughly analyzed. It
allows to investigate the dynamics and to get exact results
(soliton \cite{Kru64, Zab65, Dod82} and breather \cite{Cam96,
Sie88, Dau93, Aub94, Mac94} solutions), to analyze regular and
stochastic regimes and to find the corresponding thresholds. The
FPU preprint also initiated the investigations in the field of
``experimental mathematics'' \cite{Por09} .

About ten decades ago P.~Debye argued that the nonlinearity can be
responsible for the finite value of heat conductivity in
insulating materials \cite{Deb14}. But modern analysis shows that
it is not always the case. There are many examples where the
coefficient of heat conductivity $\varkappa$ diverges with the
increasing system size $L$ as $\varkappa \propto L^{\alpha}$ where
$\alpha > 0$, and $\varkappa \to \infty$ in the thermodynamic
limit ($L \to \infty$). Most of momentum conserving
one-dimensional nonlinear lattices with various types of
nearest-neighbor interactions have this unusual property (see,
e.g., \cite{Lep03, Cas05, Lic08}\,). Moreover, some other systems,
-- two- \cite{Bar07, Lip00, Sai10} and three-dimensional lattices
\cite{Shi08}, polyethylene chain \cite{Hen09}, carbon nanotubes
\cite{Yao05, Yu_05, Mar02, Min05, Cao04} have analogous property
-- diverging heat conductivity with the increasing size of the
system.

There were some conjectures explaining the anomalous heat
conductivity. Generally speaking, whenever the equilibrium
dynamics of a lattice can be decomposed into that of independent
``modes'' or quasi-particles, the system is expected to behave as
an ideal thermal conductor \cite{Lep05}. Thereby, the existence of
stable nonlinear excitations is expected to yield ballistic rather
than diffusive transport. At low temperatures normal modes are
phonons. At higher temperatures noninteracting ``gas'' of solitons
or/and breathers starts to play more significant role, and M.~Toda
was the first who suggested the possibility of heat transport by
solitons \cite{Tod79}.

Though analytical expressions for solitons can be derived only for
few continuum models described by partial differential equations,
Friesecke and Pego in a series of recent papers \cite{Fri99,
Fri02, Fri04a, Fri04b} made a detailed study of the existence and
stability of solitary wave solutions on discrete lattices with the
Hamiltonian $H = \sum_i \frac12 p_i^2 + u(y_i)$, where $y_i = x_i
- x_{i-1}; \ p_i = \dot x_i$. It has been proven that the systems
with this Hamiltonian and with the following generic properties of
nearest-neighbor interactions: $u'(0) = 0; \ u''(0) > 0; \ u'''(0)
\neq 0$ has a family of solitary wave solutions which in the small
amplitude, long-wavelength limit have a profile close to that of
the KdV soliton. It was also shown \cite{Hof08} that these
solutions are asymptotically stable. Thus most acceptable point of
view on the origin of anomalous heat conductivity in nonlinear
lattices is as follows: phonons are responsible for heat
conductivity at low temperatures, and at high temperatures --
solitons \cite{Li_05,Vil02}.

Rather confusing experimental and numerical results are
demonstrated in literature about the dependence of heat
conductivity on different parameters, -- model under
consideration, types of boundary conditions, used thermostat and
temperature. For instance, temperature dependence of heat
conductivity in carbon nanotubes decreases as $\varkappa \sim 1/T$
at $T > 10$ K \cite{Mar03}; experimentally is found \cite{Yu_05}
that $\varkappa$ also decreases with the growth of temperature.
Different temperature dependencies $\varkappa$ vs. $T$ were found
in 1D nonlinear lattices. For $\beta$-FPU lattice: $ \varkappa
\sim N^{\alpha} T^{-1}$ at  $T \lesssim 0.1$  and $ \varkappa \sim
N^{\alpha} T^{1/4}$ at $T > 50$  \cite{Aok01} what is usually
observed in insulating crystals. For the interparticle harmonic
potentials and on-site potentials (e.g. Klein-Gordon chains)
$\varkappa \sim T^{-1.35}$, i.e. heat conductivity decreases with
the growth of temperature \cite{Aok00}. But there exists firm
theoretical background \cite{Nar02} that the exponent $\alpha$ in
the dependence $\varkappa \propto N^{\alpha}$ is the universal
constant $\alpha \approx 1/3$ in momentum-conserving systems.

The calculation of heat conductivity at small temperature
gradients is an additional problem. Usually these calculations are
very time consuming because of great fluctuations of heat current
and statistical averaging over large number of MD trajectories is
necessary.

The paper organized as follows. In section \ref{sec:heat} the heat
conductivity is considered when the temperature gradient $\nabla
T$ is small. The explicit contribution to the heat conductivity is
extracted by the decomposing of the total process into two parts.
The first one is the equilibrium process at temperatures $T$ of
both lattice ends, and the second -- non-equilibrium, when one
lattice end has temperature $\Delta T$ and the other -- zero
temperature. Namely the latter process is responsible for the heat
conductivity. This method allows to find the threshold temperature
$T_{\rm thr}$. And though $T_{\rm thr}$ scales with the lattice
length $N$ as $T_{\rm thr} \sim N^{-3}$, unusual dynamics can be
revealed on nanoscale when both $T$ and $N$ are small. The low
temperature thermodynamics can be adequately described in terms of
harmonic lattice with temperature renormalized rigidity
coefficients.

Section \ref{sec:sol_breath} is independent of the previous one
and is aimed at elucidating the role of solitons and breathers in
the heat conductivity. Solitons and breathers are found as the
solutions of the modified Korteweg -- de~Vries equation. This
equation is obtained in the continuum limit from the discrete
$\beta$-FPU lattice. Soliton and breather solution are checked in
numerical simulation and demonstrate very high stability.

Necessary details of the derivation of accurate soliton and
breather solutions in the continuum limit are given in Appendix.

%%%%%%%%%%%%%%%%%%%%%%%%%%%%%%%%%%%%%%%%%%%%%%%%%%%%%%%%%

\section{Heat conductivity in the $\beta$-FPU lattice}
    \label{sec:heat}

We consider the one-dimensional $\beta$-FPU lattice of $N$
oscillators with the interaction of nearest neighbors
\begin{equation}
  \label{2-1}
  U = \sum\limits_i \dfrac{\alpha}{2} (x_i - x_{i-1})^2 +
                    \dfrac{\beta}{4} (x_i - x_{i-1})^4
\end{equation}
(usually the dimensionless potential will be used below, that is
$\alpha = \beta = m = 1$). Nonequilibrium conditions are necessary
for the heat transport simulation. The most abundant method is the
placement of the lattice into the heat bath with different
temperatures of left $T_+$ and right $T_-$ ends ($T_+ > T_-$).
Different types of heat reservoirs are thoroughly analyzed in
\cite{Lep03}. The usage of the Langevin forces with the noise
terms and friction forces acting on the left $F_+ = \xi_+ - \gamma
\dot x_1$ and right  $F_- = \xi_- \gamma \dot x_N$ oscillators is
the common practice ($\gamma = 1$ is also put for brevity).
$\left\{ \xi_{\pm} \right\}$ are independent Wiener processes with
zero mean and the correlator $\left< \xi_{\pm}(t_1) \,
\xi_{\pm}(t_2) \right> = 2 T_{\pm} \, \delta(t_1 - t_2)$. $\Delta
T = (T_+ - T_-)$ is the temperature difference. The generalized
Langevin dynamics with a memory kernel and colored noises is also
suggested \cite{Wan07} to correctly account for the effect of the
heat baths.

The following set of stochastic differential equations (SDEs)
\begin{equation}
  \label{2-2}
   \ddot x _i = - \dfrac{\partial U}{\partial x_i} + \delta_{i1}
   F_+ + \delta_{iN} F_-
\end{equation}
is usually solved to find the heat flux $J$. And the local heat
flux (power transmitted from $i$th to $(i+1)$th oscillator) is
\cite{Zha02}
\begin{equation}
  \label{2-4}
  J_{i \to i+1} =  F_{i \to i+1}\, \dot x_{i+1}  ;
  \qquad F_{i \to i+1} \equiv -U'(x_{i+1} - x_i),
\end{equation}
where $F_{i \to i+1}$ is a shorthand notation for the force
exerted by the $i$th on the $(i+1)$th oscillator. The total heat
flux $J$ can be found as the mean $J = (N-1)^{-1} \sum_i^{N-1}
J_{i \to i+1}$.

%%%%%%%%%%%%%%%%%%%%%%%%%%%%%%%%%%%%%%%%%%%%%%%%%%%%%%%%%

\subsection{Equilibrium and non-equilibrium contributions
            to the heat conductivity}

\begin{figure}
\begin{center}
\includegraphics[width=100 mm,angle=0]{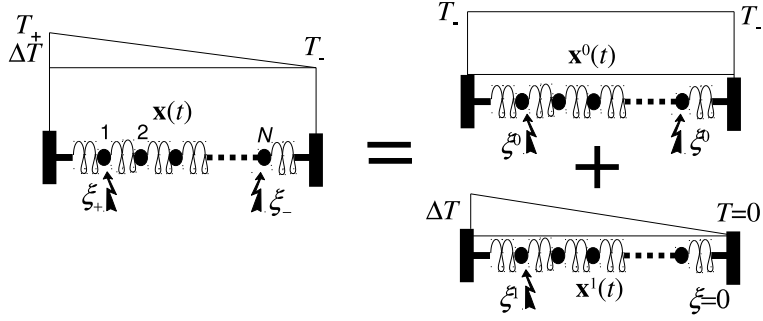}
\end{center}
 \caption{
  \label{fig_01}
  Schematic representation of the total
  process ${\bf x}(t)$ as sum of equilibrium ${\bf x}^0(t)$ and
  non-equilibrium ${\bf x}^1(t)$ processes.
 }
\end{figure}

If $T_- \neq 0$ then the process of heat conductivity can be
formally decomposed into two contributions: the first one --
equilibrium process with equal temperatures $T_-$ of both lattice
ends; and the second -- nonequilibrium process with temperature
$\Delta T$ of the left lattice end and zero temperature of the
right end (see Fig.~\ref{fig_01}) (by `process' we hereafter
assume for brevity the solution $ {\bf x} (t) =  \left\{ x_1(t),
x_2(t), \ldots , x_N(t) \right\}; \, {\bf v} (t) =  \left\{
v_1(t), v_2(t), \ldots , v_N(t) \right\} $ of the corresponding
SDEs).

Namely the second process is responsible for the heat transport
taking place on(?) the background of equilibrium process. Once
this approach is utilized then the noise terms in \eqref{2-2},
owing to their independence, are $\left\{ \xi_+ \right\} = \left\{
\xi^0 \right\} + \left\{ \xi^1 \right\} $ and $ \left\{ \xi_-
\right\} = \left\{ \xi^0  \right\} $ for the left and right
lattice ends, correspondingly  Superscripts `0' and `1' refer to
equilibrium and nonequilibrium processes. The total process ${\bf
x}(t) $ can be represented as the sum
\begin{equation}
  \label{2-5}
    {\bf x}(t) = {\bf x}^0(t) + {\bf x}^1(t),
\end{equation}
where ${\bf x}^0(t)$ is the equilibrium (Gibbs's) process at
temperature $T_-$, and ${\bf x}^1(t)$ -- nonequilibrium,
responsible for the energy transport, process. The corresponding
stochastic dynamics is
\begin{equation}
  \label{2-6}
  \hspace{-1.3 cm} \ddot x_i^0 =  - \dfrac{\partial U^0}{\partial x_i^0} +
                   \delta_{i1} (\xi^0 - \dot x^0_1) +
                   \delta_{iN} (\xi^0 - \dot x^0_{N}),
\end{equation}
\begin{equation}
  \label{2-7}
   \ddot x_i^1 =   - \left[
                  \dfrac{\partial U}{\partial x_i} -
                  \dfrac{\partial U^0}{\partial x_i^0}
                     \right] +
                  \delta_{i1} (\xi^1 - \dot x^1_1) +
                  \delta_{iN} (- \dot x^1_N),
\end{equation}
and the sum of equations \eqref{2-6} and \eqref{2-7} is identical
to the parent equation \eqref{2-2}. Random values $\left\{ \xi^0
\right\}$ and $\left\{ \xi^1 \right\}$ obey the identities $\left<
\xi^0(t_1) \xi^0(t_1) \right> = 2  T_- \delta(t_1 - t_2)$ and
$\left< \xi^1(t_1) \xi^1(t_1) \right> = 2 \, \Delta T \delta(t_1 -
t_2)$; $U^0$ is the total energy \eqref{2-1} where the arguments $
x_1(t), x_2(t), \ldots, x_N(t) $ of the total process are replaced
by coordinates of the equilibrium process $x_1^0(t), x_2^0(t),
\ldots, x_N^0(t)$. Expression in the square brackets in
\eqref{2-7} is the difference of forces acting on the $i$th
oscillator from the total process ${\bf x}(t)$ and equilibrium
process ${\bf x}^0(t)$.  It is worth mentioning that this force is
the random value, and the process ${\bf x}^1(t)$ (heat transport)
is realized in the lattice with {\it time-dependent random
potentials}. The problem of heat conductivity in the random
time-independent potentials was analyzed in \cite{Joh08}.

\begin{figure}
\begin{center}
\includegraphics[width=80mm,angle=0]{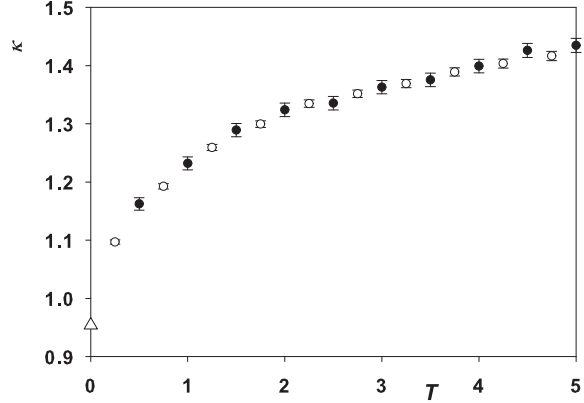}
\end{center}
 \caption{
  \label{fig_02}
  Temperature dependence of heat conductivity for the lattice of $N=5$
  oscillators. Filled circles: solution of standard SDEs \eqref{2-2};
  empty circles: SDEs
  \eqref{2-6}-\eqref{2-7}. Averaging over $100$ MD trajectories
  $10^4$ time units (t.u.) each. $T_- = 0{.}2$,
  $\Delta T = 0{.}01 T_-$. Triangle up at
  $T = 0$ is the exact value in the harmonic approximation ($\beta = 0$).
         }
\end{figure}

Equation \eqref{2-6} describes the system embedded in the heat
reservoir at temperature $T_-$ of both lattice ends. And ${\bf
x}^0(t)$ is the stationary equilibrium process described by the
canonical Gibbs distribution. Process  ${\bf x}^1(t)$ is
responsible for the heat transport and the Wiener  process
$\left\{ \xi^1(t) \right\}$ on the left lattice end defines
temperature $\Delta T$. Right lattice end has zero temperature
(only the friction force acts on this oscillator). An expression
for the local heat flux is
\begin{equation}
  \label{2-8}
 J_{i \to i+1} =
        \left[ F_{i \to i+1}({\bf x}) - F_{i \to i+1}({\bf x}^0) \right]\, \dot
        x_{i+1}^0 ,
\end{equation}
and the equilibrium process ${\bf x}^0$ does not transfer energy:
$\left<  F_{i \to i+1}({\bf x}^0) \, \dot x^0_{i+1} \right> \equiv
0$, where $\left< \ldots \right>$ stands for the time average. It
is essential that the heat flux \eqref{2-8} is the small
difference of large values from processes ${\bf x}(t)$ and ${\bf
x}^0(t)$. This is the reason why MD simulation gives large
fluctuation when $\Delta T \to 0$ and $T$ is not low. It can be
shown that the time of computation increases $\propto (\Delta
T)^{-2}$ if the standard error is fixed. The comparison of two
approaches (solving of standard SDEs \eqref{2-2} and
\eqref{2-6}-\eqref{2-7}) is shown in Fig.~\ref{fig_02} and results
coincide with very good accuracy.

Some results in this paper are obtained for the number of
oscillators $N = 5$. It may appear that this value is too small.
For instance, ``standard'' simulations require up to $\sim 10^4$
particles and $\sim 10^8$ integration steps plus ensemble
averaging \cite{Lep03}. But our results are aimed at establishing
some new issues where number of particles is unessential. Lattices
with larger number of oscillators were tested when necessary.

Relative displacements $(x_i^0 - x_{i-1}^0)$ and $(x_i^1 -
x_{i-1}^1)$ for processes ${\bf x}^0(t)$ and ${\bf x}^1(t)$ are
shown in Fig.~3. These values characterize the energy fluxes in
the lattice. Energy fluxes to the left and to the right are equal
on average for process ${\bf x}^0(t)$ as it is the equilibrium
process without energy transfer. But the energy flux is directed
mainly to the right for process ${\bf x}^1(t)$ (right panel). Low
temperature is chosen for the better illustration.
\begin{figure}
  \begin{center}
\includegraphics[width=60mm,angle=0]{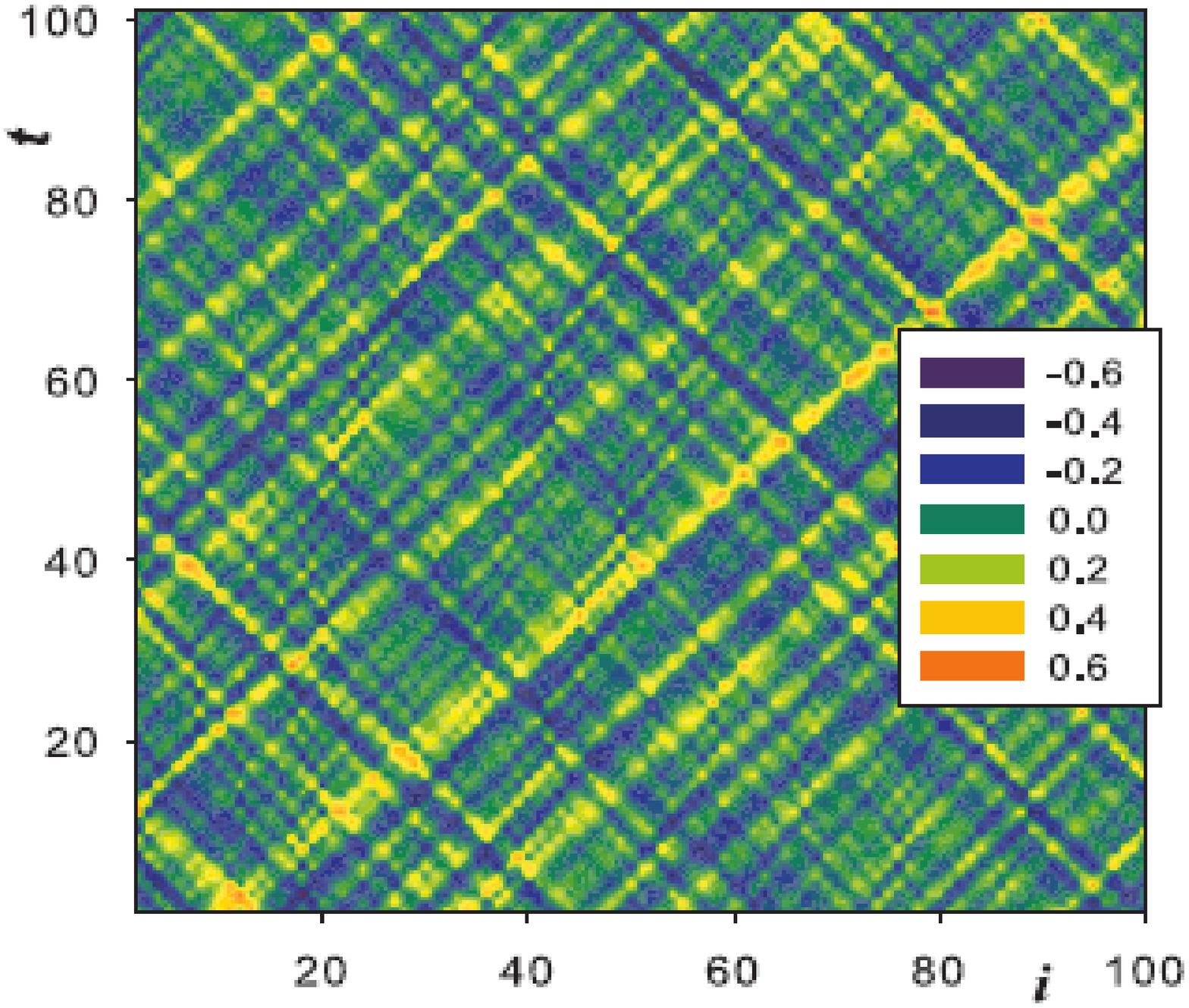}
\includegraphics[width=68mm,angle=0]{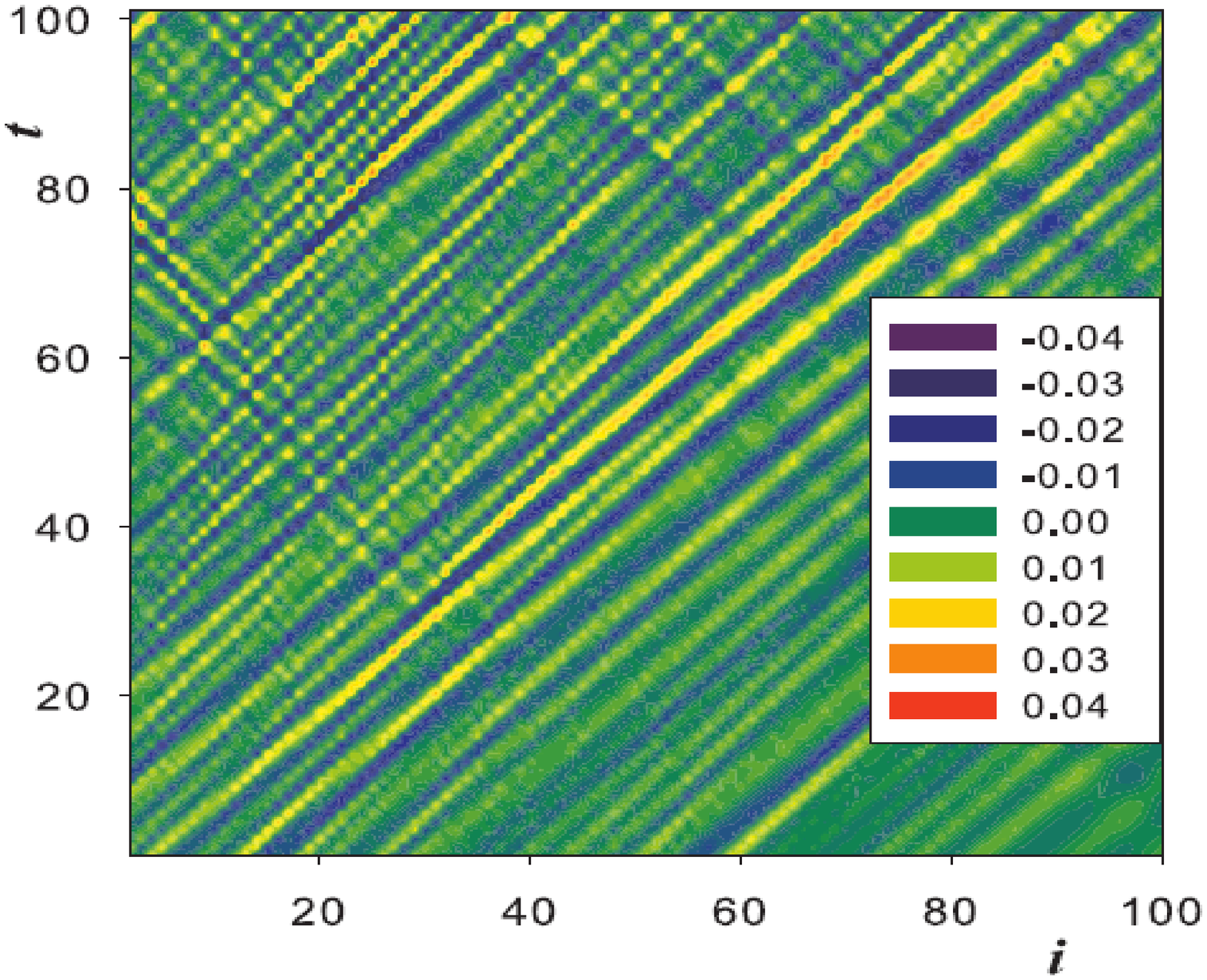}
  \end{center}
 \caption{
  \label{fig_3}
  Spatiotemporal evolutions of relative displacements $(x_i^0 - x_{i-1}^0)$
  and $(x_i^1 - x_{i-1}^1)$ for the equilibrium
  ${\bf x}^0(t)$ (left panel) and nonequilibrium ${\bf x}^1(t)$ (right panel) processes.
  $N = 100$, $T_- = 0{.}05$, $\Delta T = 0{.}0001$.
         }
\end{figure}

\begin{figure}
\begin{center}
\includegraphics[width=80mm,angle=0]{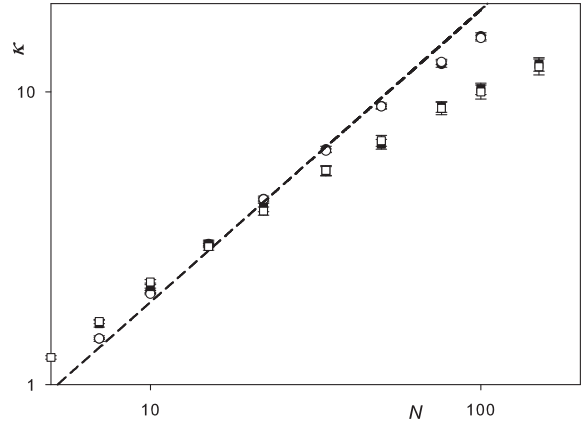}
\end{center}
 \caption{
  \label{fig_4}
  Coefficient of heat conductivity for the $\beta$-FPU lattice for
  $N = 7-150$ oscillators. Squares: $T_-=1$, circles:
  $T_-=0{.}1$. Filled symbols -- results obtained by the solution
  of standard SDEs \eqref{2-2}, empty symbols -- SDEs
  \eqref{2-6}-\eqref{2-7}. Averaging over 200 MD trajectories
  $3\,10^4$ t.u. $\Delta T = 0{.}01T_-$. Dashed line -- harmonic
  approximation. Filled symbols are practically fully covered by
  empty symbols and are invisible.
 }
\end{figure}

The dependence of heat conductivity on the oscillators number $N$
is shown in Fig.~\ref{fig_4} at two value of temperature $T_-$.
Results coincide with very good accuracy. Inharmonicity becomes
negligible at low temperature and heat conductivity at $T_- =
0{.}1$ (cirles in Fig.~\ref{fig_4}) coincides with the heat
conductivity of the harmonic lattice (dashed line) with good
accuracy. The analytical solution of the heat conductivity for the
harmonic lattice is given in \cite{Rie67}.
There should be solved twice as large SDEs \eqref{2-6}-\eqref{2-7}
in suggested approach as that in standard scheme \eqref{2-2}. But
this approach has some undoubted merits as discussed below.

\subsection{Heat conductivity at small temperature gradients}

One of the goals of the present paper is the computation of heat
conductivity at very small temperature gradients. With this in
mind an expression for the heat flux is analyzed in more details.
The expression for the local heat flux \eqref{2-8} can be
rewritten as
\begin{equation}
  \label{2-9}
  \begin{split}
   J_{i \to i+1} =   & \left[ F_{i \to i+1}({\bf x}) -
                                      F_{i \to i+1}({\bf x}^0)  \right]
    \dot x_{i+1}^1  = \\
 {} &    \left[ -(x_{i+1} - x_i) -(x_{i+1} - x_i)^3 +
  (x^0_{i+1} - x^0_i) + (x^0_{i+1} - x^0_i)^3   \right]
                       \dot x_{i+1}^1 ,
  \end{split}
\end{equation}
where $\dot x_{i+1}^1$ -- velocity of $(i+1)$th oscillator in
process ${\bf x}^1$. Taking in mind that $x_i = x_i^0 + x_i^1$,
the expression \eqref{2-9} can be transformed to
\begin{equation}
  \label{2-90}
  J_{i \to i+1} =  \left\{ - (x_{i+1}^1 - x_i^1)
                     \left[ 1 + 3 \, (x_{i+1}^0 - x_i^0)^2
                     \right] +
                 (x_{i+1}^1 - x_i^1)^2
                     \left[1 + 3 \, (x_{i+1}^0 - x_i^0)
                     \right]
                  \right\}  \dot x_{i+1}^1 .
\end{equation}

Processes ${\bf x}^0(t)$ and ${\bf x}^1(t)$ have different ranges
of specific energies.  Noise terms $\left\{ \xi^1 \right\}$, which
provide temperature $\Delta T$, are of the order $ \xi^1 \sim
\sqrt{\Delta T}$ (as $\left< \xi^1(t_1) \xi^1(t_2) \right> \sim
\Delta T$). And one can expect that process ${\bf x}^1(t)$ has the
same order ${\bf x}^1(t) \sim \sqrt{\Delta T}$ because equations
\eqref{2-7} become linear in the limit $\Delta T \to 0$ when
$\xi^1 \to 0$. Expression in curly brackets in \eqref{2-90} is the
polynomial of the third degree in the square root of temperature
difference $\sqrt{\Delta T}$. Taking into account that the
velocity $\dot x_{i+1}^1$ is also of the order $\sim \sqrt{\Delta
T}$, expression \eqref{2-90} is the polynomial of the forth degree
in $\sqrt{\Delta T}$. But the coefficient of heat conductivity is
determined by the relation $J/\Delta T$. Then terms of the third
and forth orders can be neglected at $\Delta T \to 0$. Then
\eqref{2-90} is simplified to
\begin{equation}
  \label{2-91}
     J_{i \to i+1} =  - \left( x_{i+1}^1 - x_i^1 \right)
     \left[ 1 + 3 \, \left( x_{i+1}^0 - x_i^0 \right)^2   \right]
     \dot x_{i+1}^1 .
\end{equation}

An expression for the potential energy, corresponding to process
${\bf x}^1(t)$, can be derived analogously. This energy is the
difference of potential energies $U({\bf x}) - U^0({\bf x}^0)$ and
again, using coordinates ${\bf x}$ and ${\bf x}^0$, and preserving
only terms quadratic in $(x_{i+1}^1 - x_{i}^1)$, one can get the
potential energy for process ${\bf x}^1$ in the form
\begin{equation}
  \label{2-11}
  U^1 = \dfrac12 \sum_i g_{i+1}(t) \, (x_{i+1}^1 - x_{i}^1)^2,
  \qquad
  g_{i+1}(t) = 1 + 3 \, [x_{i+1}^0(t) - x_{i}^0(t)]^2 \,,
\end{equation}
where  $g_i(t)$ are {\it time-dependent random} coefficients of
rigidities determined by process ${\bf x}^0(t)$.

It is illuminating to note that in the limit $\Delta T \to 0$ the
problem of heat conductivity in the $\beta$-FPU lattice is reduced
to the {\it harmonic} lattice with random coefficients.
Corresponding SDEs have noise terms with friction forces on the
left oscillator and zero temperature (only viscous forces) on the
right oscillator:
\begin{equation}
  \label{2-12}
  \ddot x_i^1 =  - g_i (x_i^1 - x_{i-1}^1) + g_{i+1} (x_{i+1}^1 - x_{i}^1)  +
  \delta_{i1} (\xi^1 - \dot x_1^1) - \delta_{i N} \dot x_{N}^1
\end{equation}
and $g_i, g_{i+1}$ are defined in \eqref{2-11}. If the 1D lattice
with an arbitrary interaction potential is analyzed then the
corresponding equations are the same with rigidities
$g_i=U''(x^0_i - x^0_{i-1}$) where $U$ is the potential energy.
SDE for the system with arbitrary neighbor radius of interaction
can be written in the general form as
\begin{equation}
  \label{2-13}
  \ddot x_i^1 = - \sum_{j=1}^M {\bf \Lambda}_{ij}^0 \, x_j^1  +
  \delta_{i1} (\xi^1 - \dot x_1^1) - \delta_{i N} \dot x_{N}^1
  \,,
\end{equation}
where ${\bf \Lambda}_{ij}^0$ -- matrix of second derivatives of
potential energy depending on ${\bf x}^0$, and $M$ is the number
of neighbors.

%%%%%%%%%%%%%%%%%%%%%%%%%%%%%%%%%%%%%%%%%%%%%%%%%%%%

\subsection{Unusual dynamics of process ${\bf  x}^1(t)$ at high
temperatures}

\begin{figure}
\begin{center}
\includegraphics[width=90mm,angle=0]{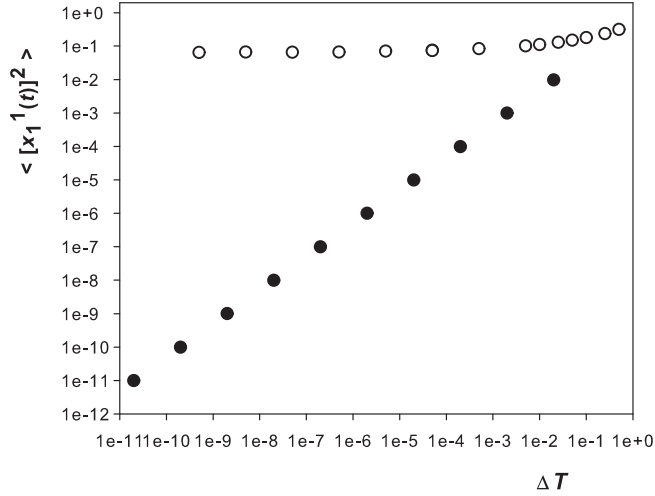}
\end{center}
 \caption{
  \label{fig_5}
Dependence of correlator $\left< (x_1^1)^2   \right>$ on the
temperature difference $\Delta T$. Filled circles:   $T = 0{.}2$,
empty circles: $T=5$. Asymptotic value $\left< (x_1^1)^2
\right>_{\Delta T \to 0} \approx 0{.}064$ at $T = 5$ (coefficient
of linear regression $0{.}9993$). Averaging over 100 MD
trajectories $10^4$ t.u. each. $N=5$. The range of $\Delta T$:
$10^{-11} \leq \Delta T/T \leq 2 \! \cdot \! 10^{-1}$.
 }
\end{figure}

The process ${\bf  x}^1(t)$ can be characterized by some time
average correlators. The correlator $\left< \left[ x_1^1(t)
\right]^2 \right>$ was analyzed for the better understanding of
the dynamics of process ${\bf x}^1(t)$. One can expect that
$\left< \left[ x_1^1(t) \right]^2 \right> \sim \Delta T$ as
discussed above. Two temperatures of the background process $T_-$
were tested: $T_1 = 0{.}2$ and $T_2 = 5$ (hereafter subindex `--'
is omitted, that is $T_- = T$). Results are shown in
Fig.~\ref{fig_5}.

As expected, the correlator $\left< \, [x_1^1(t)]^2 \right>$
linearly depends on $\Delta T$: $\left< \, [x_1^1(t)]^2 \right>
\sim \Delta T$ at $T_1 = 0{.}2$. But the case is quite different
at $T_2 = 5$: $\left< \, [x_1^1(t)]^2 \right>$ reaches the
stationary value $\approx 0{.}064$ in the limit $\Delta T \to 0$.
It means that there exists some undamped stationary process ${\bf
x}^1(t)$ at high temperatures $T$ of the background process ${\bf
x}^0(t)$ even in the limit $\Delta T \to 0$. These results also
imply an existence of a threshold temperature $T_{\rm thr}$
separating two regimes -- damped at low temperatures and undamped
at high temperatures.

%%%%%%%%%%%%%%%%%%%%%%%%%%%%%%%%%%%%%%%%%%%%%%%%%%%%%%

\subsection{Threshold temperature}

Any process ${\bf x}^1(t)$ damps out at low temperatures and
flattens out to a stationary value at higher temperatures even in
the limit $\Delta T \to 0$, and the temperature $T$ of process
${\bf x}^0(t)$ defines  different damping rates. Bearing this in
mind, it is convenient to excite some auxiliary process
$\widetilde{\bf x}^1(t)$ over the background process ${\bf
x}^0(t)$ and to analyze it.

Coordinates and velocities of process $\widetilde{\bf x}^1(t)$ get
random increments $\frac12 \sum_i [{\widetilde x}_i^1(t=0)]^2 +
\frac12 \sum_i [ {\widetilde v}_i^1(t=0)]^2 = 0{.}5$. The
particular choice of initial conditions does not influence the
final results. The total dynamics is the sum of two processes
${\bf x}(t) = {\bf x}^0(t) + \widetilde{\bf x}^1(t)$.

Stochastic differential equations for the process ${\bf \widetilde
x}^1(t)$ are
\begin{equation}
  \label{2-10}
   \ddot {\widetilde x}_i^1 =   - \left[
                  \dfrac{\partial  U } {\partial x_i} -
                  \dfrac{\partial  U^0 }{\partial x_i^0}
                     \right] -
                  \delta_{i1}\dot {\widetilde x}^1_1 -
                  \delta_{iN} \dot {\widetilde x}^1_{N}
\end{equation}
and only viscous forces acts at the extreme left and right
oscillators. $U$ and $U^0$ are potential energies with coordinates
${\bf x}(t)$ and ${\bf x}^0(t)$, correspondingly. Stochastic
dynamics \eqref{2-10} is implicitly ruled out by the temperature
$T$ of process ${\bf x}^0(t)$.

\begin{figure}
!\begin{center}
\includegraphics[width = 75mm, angle=0]{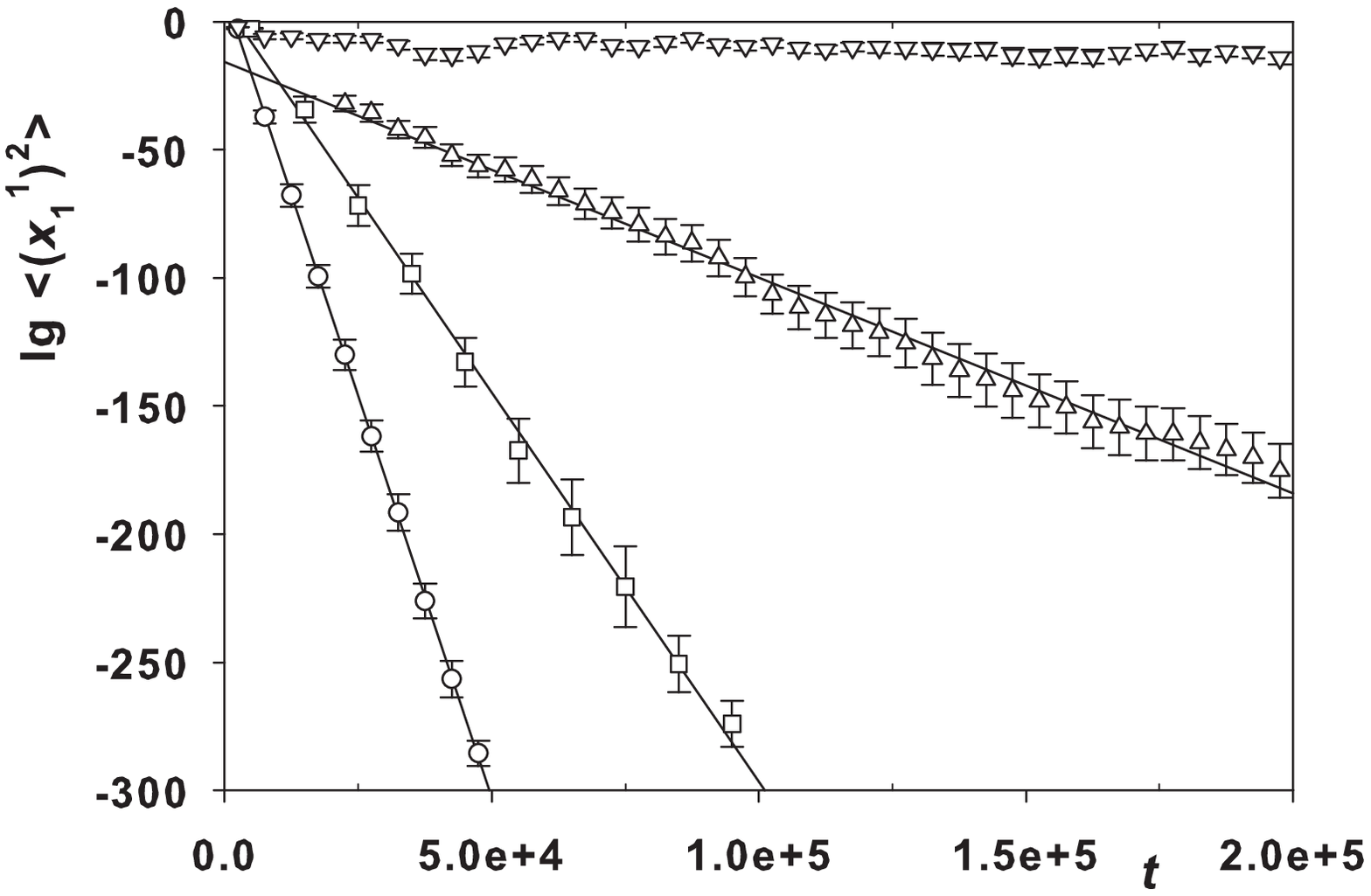}
\includegraphics[width = 70mm, angle=0]{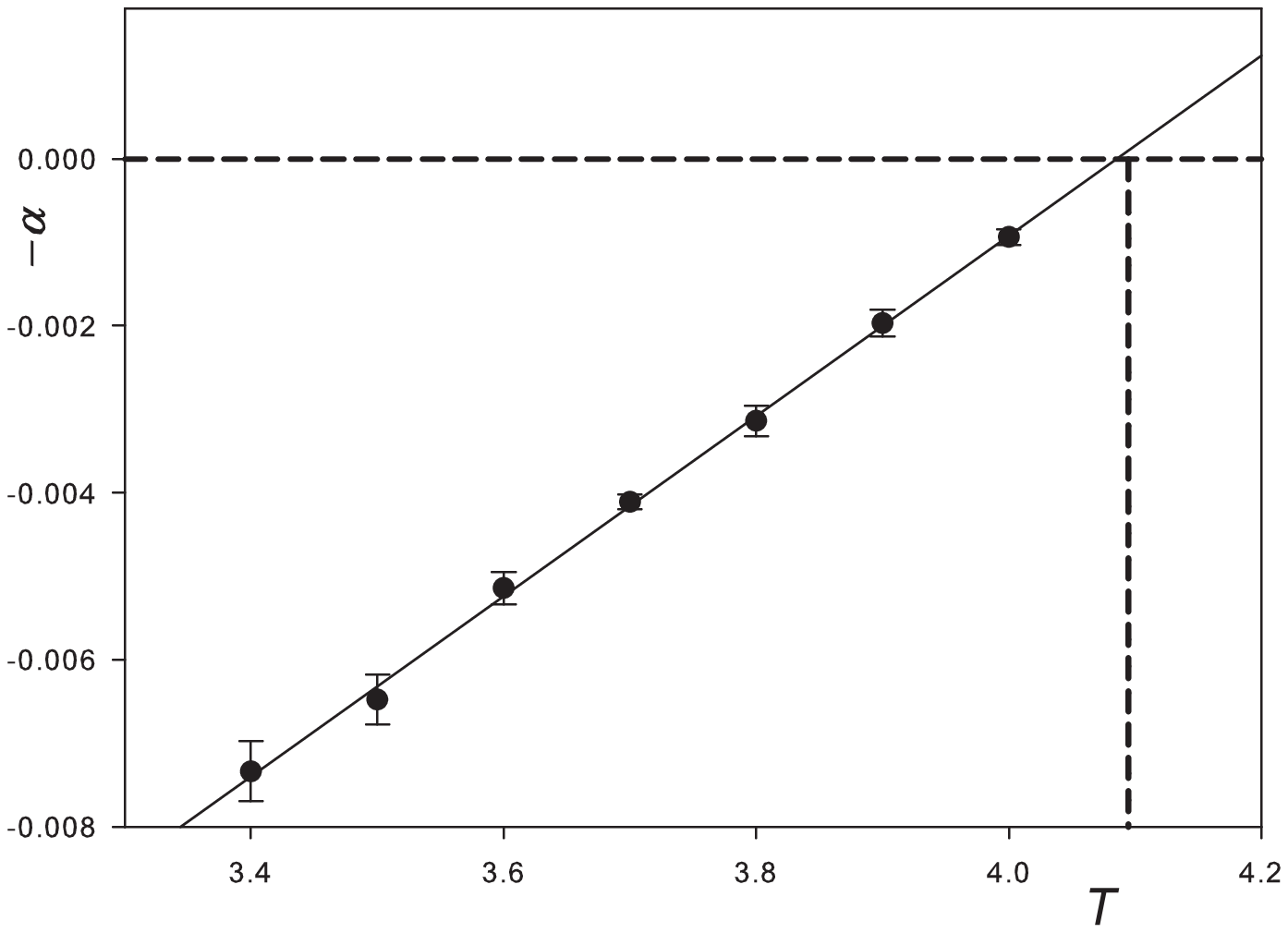}
!\end{center}
 \caption{
  \label{fig_6}
  a) Exponential damping of process ${\bf \widetilde x}^1(t)$ at
  different temperatures: $T=3{.}5$ (circles), $T=3{.}8$ (squares),
  $T=4{.}0$ (triangles up), $T=4{.}2$ (triangles down). Solid
  lines -- linear regressions. Averaging time  $\sim \! 5\,000-10\,000$
  t.u. 20 trajectories ${\bf x}^0$ were used to estimate the standard
  error. b) Damping coefficient $(-\alpha)$ as the function of temperature
  $T$ of process ${\bf x}^0(t)$. Damping stops $(\alpha = 0)$ at
  $T_{\rm thr} \simeq 4{.}07$. $N=5$.
 }
\end{figure}

To find the threshold temperature we initially consider the case
of small temperature $T$ when process ${\bf \widetilde x}^1(t)$ is
damped out. Its damping is determined by the viscous friction of
left $(-\dot {\widetilde x}^1_1)$ and right $(-\dot {\widetilde
x}^1_{N})$ oscillators in \eqref{2-10}. Gradually increasing the
temperature its threshold value can be found when process ${\bf
\widetilde x}^1(t)$ becomes undamped. The damping of mean squared
displacement of the first oscillator $[{\widetilde x}_1^1(t)]^2$
was calculated. Process $\widetilde{\bf x}^1(t)$ exponentially
decays $\left< \, [\widetilde x_1^1(t)]^2 \right> \propto
\exp(-\alpha t)$ and $\alpha$ depends on $T$ (see
Fig.~\ref{fig_6}a).
The temperature dependence of coefficient $\alpha$ is shown in
Fig.~\ref{fig_6}b and $T_{\rm thr} \simeq 4{.}07$ when $\alpha =
0$.

\begin{figure}
\begin{center}
\includegraphics[width=80mm,angle=0]{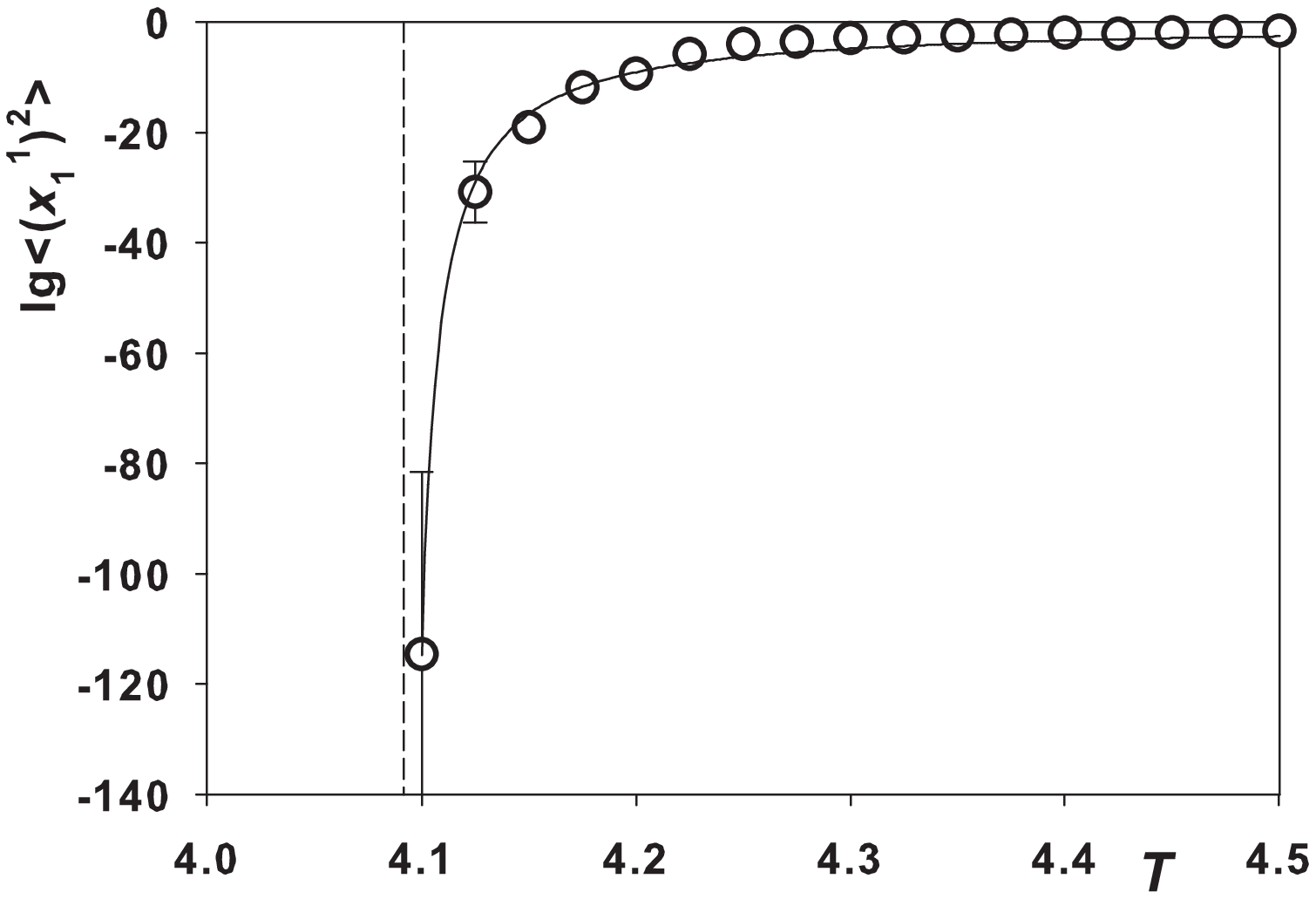}
\end{center}
 \caption{
  \label{fig_7}
  Stationary values $\left< (\widetilde x_1^1)^2 \right>$ at
  $T > T_{\rm thr}$.
  Time of averaging $~10^6$ t.u. The temperature dependence is
  approximated by the function
  $\left< [\widetilde x_1^1(t)]^2 \right>
  \sim \exp[-b /(T-T_{\rm thr})]$ (solid line). $N=5$.
 }
\end{figure}

Next method to find the threshold temperature is moving ``from up
to down'', going from higher to lower temperatures. At high
temperatures there exists the stationary process arising from
random forces $\Phi_i = \left[ {\partial  U }/ {\partial x_i} -
{\partial  U^0 }/{\partial x_i^0} \right]$ (see~\eqref{2-10}). And
process ${\bf \widetilde x}^1(t)$ decreases in a sense that all
quadratic mean values tend to zero as temperatures approaches
$T_{\rm thr}$ from above. When the temperature reaches its
threshold value, process ${\bf \tilde x}^1(t)$ is totally damped
(see Fig.~\ref{fig_7}). The found threshold temperature is $T_{\rm
thr} \simeq 4{.}09$.

Strange behavior of process ${\bf x}^1(t)$ is basically explained
by {\it time-dependent random} forces $\Phi_i = \left[ {\partial
U}/{\partial x_i} - {\partial U^0}/{\partial x_i^0} \right]$ (see
\eqref{2-7}) rather then random Langevin forces $\xi^1 \sim \sqrt{
\Delta T}$. And the plateau for the correlator $\left<
[x^1_1(t)]^2 \right> \approx 0{.}064$ at $\Delta T \to 0$ is
determined exclusively by the background process ${\bf x}^0$.

\begin{figure}
\begin{center}
\includegraphics[width=90mm,angle=0]{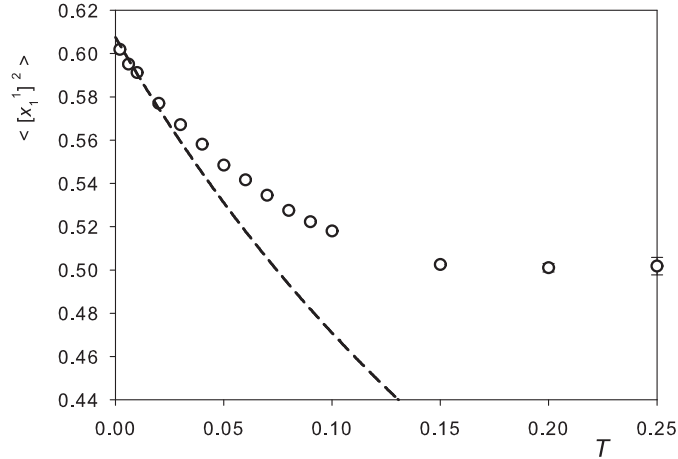}
\end{center}
 \caption{
  \label{fig_8}
  Temperature dependence of the mean squared displacement
  $\left< [x_1(t)]^2 \right>$. Circles -- MD simulation of SDEs
  \eqref{2-12}; dashed line -- model of mean rigidities in the harmonic
  approximation. Averaging over 20 trajectories $2 \, 10^4$ t.u.
  each.
  }
\end{figure}

This conjecture can be additionally supported. Let us consider the
case of low temperatures $T$ when process ${\bf x}^0$ is ``weak''.
Then the rigidity coefficients $g_i$ in \eqref{2-11} are close to
unity. The lattice where actual rigidity coefficients $g_i$
\eqref{2-11} are substituted by the mean values taken from the
equilibrium Gibbs distribution: $\left< g_i \right> = g_0(T)$ and
$g_0(T) = 1 + 3\left< (x^0_i -x^0_{i-1})^2 \right>$ is considered
as an example. This harmonic model is exactly solvable and results
are shown in Fig.~\ref{fig_8}

One can see that process $\left< [x_1^1(t)]^2 \right>$ damps out
in the model with constant rigidity in contrast to the case when
actual values \eqref{2-11} are used. And the growth of process
$\left< [x_1^1(t)]^2 \right>$, when temperature increases, is
mainly governed by {\it fluctuations} rather then the increasing
rigidities.

Additional evidence of threshold phenomena is the one-dimensional
analogue of the Mathieu equation
\begin{equation}
  \label{Math_1}
  \ddot x = -[1 + g \cos^2(t)] \, x.
\end{equation}
Different types of solutions depend on the parameter $g$ and
initial conditions. There exists such critical value $g_{\rm cr}$
that the solution is the superposition of periodic functions at $g
< g_{\rm cr}$, and the solution diverges $\propto \exp (\pm \mu
t)$ at $g > g_{\rm cr}$.

We consider an equation for the harmonic oscillator for one
variable $x$ with friction force
\begin{equation}
  \label{Math_2}
  \ddot x = - k(t) x - \dot x,
\end{equation}
where $k(t)$ -- stochastic rigidity. This equation is similar to
equation \eqref{2-12} for process ${\bf x}^1(t)$ if $k(t)= 1 + 3
z^2(t)$ and $z(t)$ is the stochastic process generated by the
dynamics of harmonic oscillator with noise term and friction force
at temperature $T$:
\begin{equation}
  \label{Math_3}
  \ddot z = - z + \xi - \dot z
\end{equation}
and spectral property $\left<  \xi(t_1) \, \xi(t_2) \right> = 2 T
\delta(t_1 - t_2)$. The substitution $x \exp(- t/2) \to X$
excludes damping and \eqref{Math_2} becomes
\begin{equation}
  \label{Math_4}
  \ddot X = - [1 + 3 z^2(t)] X,
\end{equation}
what is similar to the Mathieu's equation \eqref{Math_1}.

Eqs. \eqref{Math_3}-\eqref{Math_4} have rich family of solutions
depending on initial conditions, temperature $T$ and the sequence
$\left\{  {\bf \xi} \right\}$. The solution is nearly harmonic
function at very low temperatures. The superposition of harmonic
functions is the solution at higher temperatures. At last there
exists such threshold temperature $T_{\rm thr}$ that the solution
diverges and is the product of harmonic functions by $\exp(\mu
t)$. The solution is the product of some stochastic process by
$\exp(\nu t)$ at much higher temperatures, and $\nu > \mu$. There
are many interesting intermediate solutions, and this problem
merits more attention. Considered examples show that an existence
of threshold phenomena is not exceptional and can occur in
different dynamical systems.

\begin{figure}
\begin{center}
\includegraphics[width=90mm,angle=0]{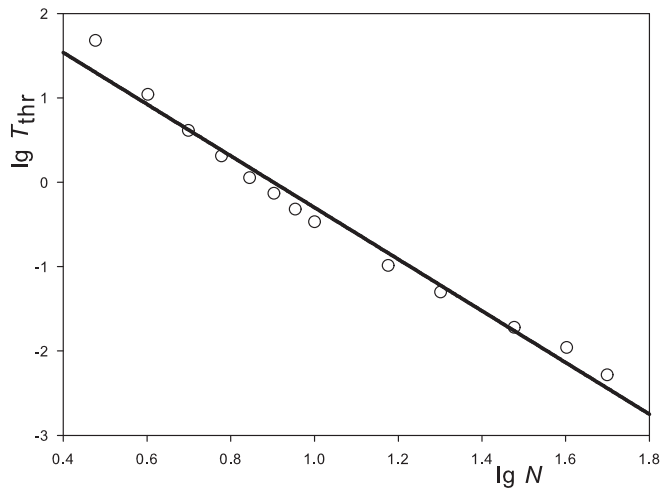}
\end{center}
 \caption{
  \label{fig_9}
  Dependence of $T_{\rm thr}$ vs. lattice length $N$ in log-log
  coordinates. Solid line is the fitting $T_{\rm thr} \sim
  N^{-3}$.
 }
\end{figure}

The threshold temperature $T_{\rm thr} \approx 4{.}1$ was found
for the lattice length $N = 5$. Larger lattice lengths were
considered and the dependence of $T_{\rm thr}$ on the lattice
length $N$ is shown in Fig.~\ref{fig_9}. Fitting gives dependence
$T_{\rm thr} \approx 6 \cdot 10^2 \, N^{-3}$.

\subsection{Time-resolved dynamics of process ${\widetilde{\bf x}}^1(t)$ }

\begin{figure}
\begin{center}
\includegraphics[width=60mm,angle=0]{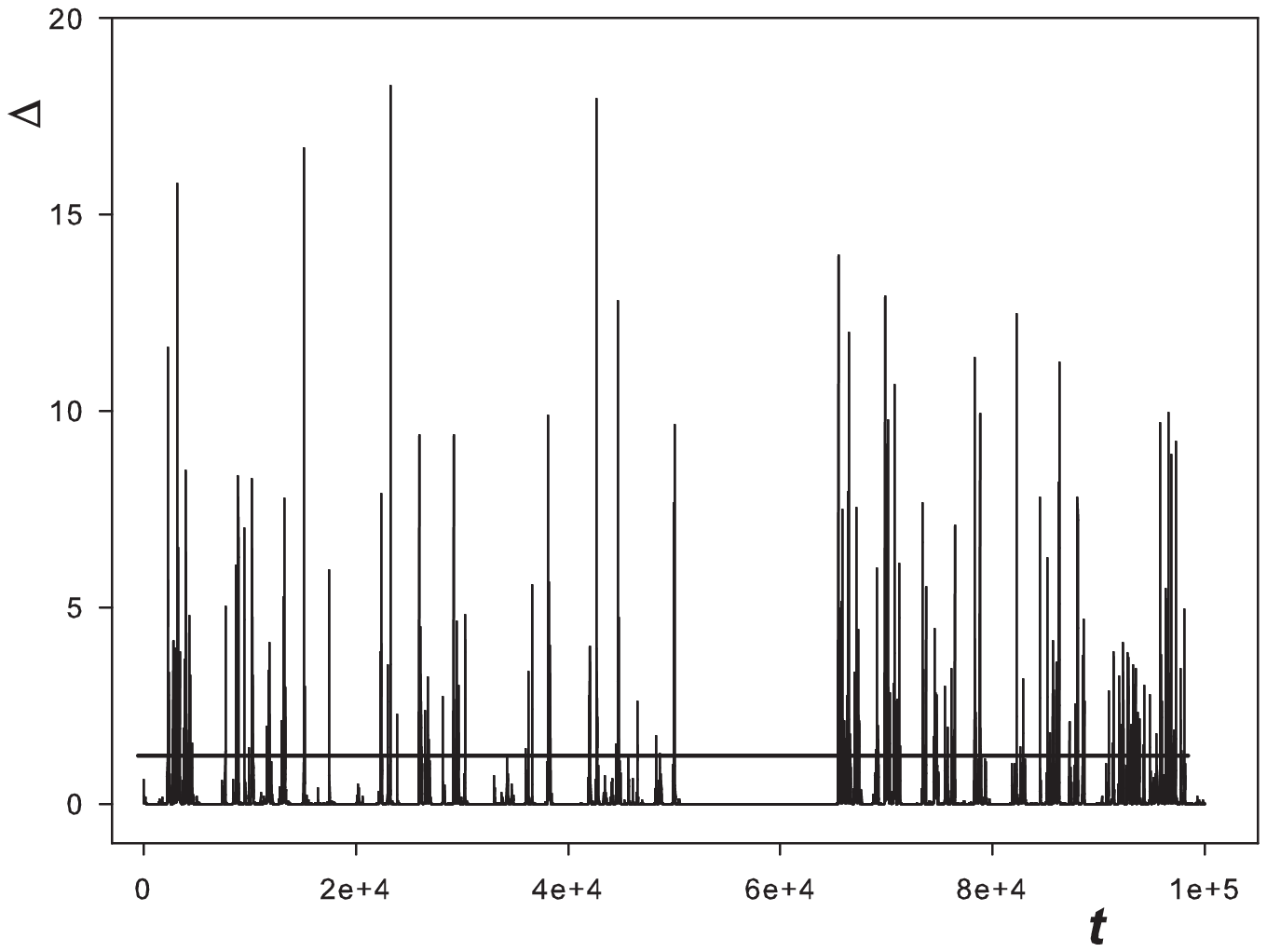}
\includegraphics[width=60mm,angle=0]{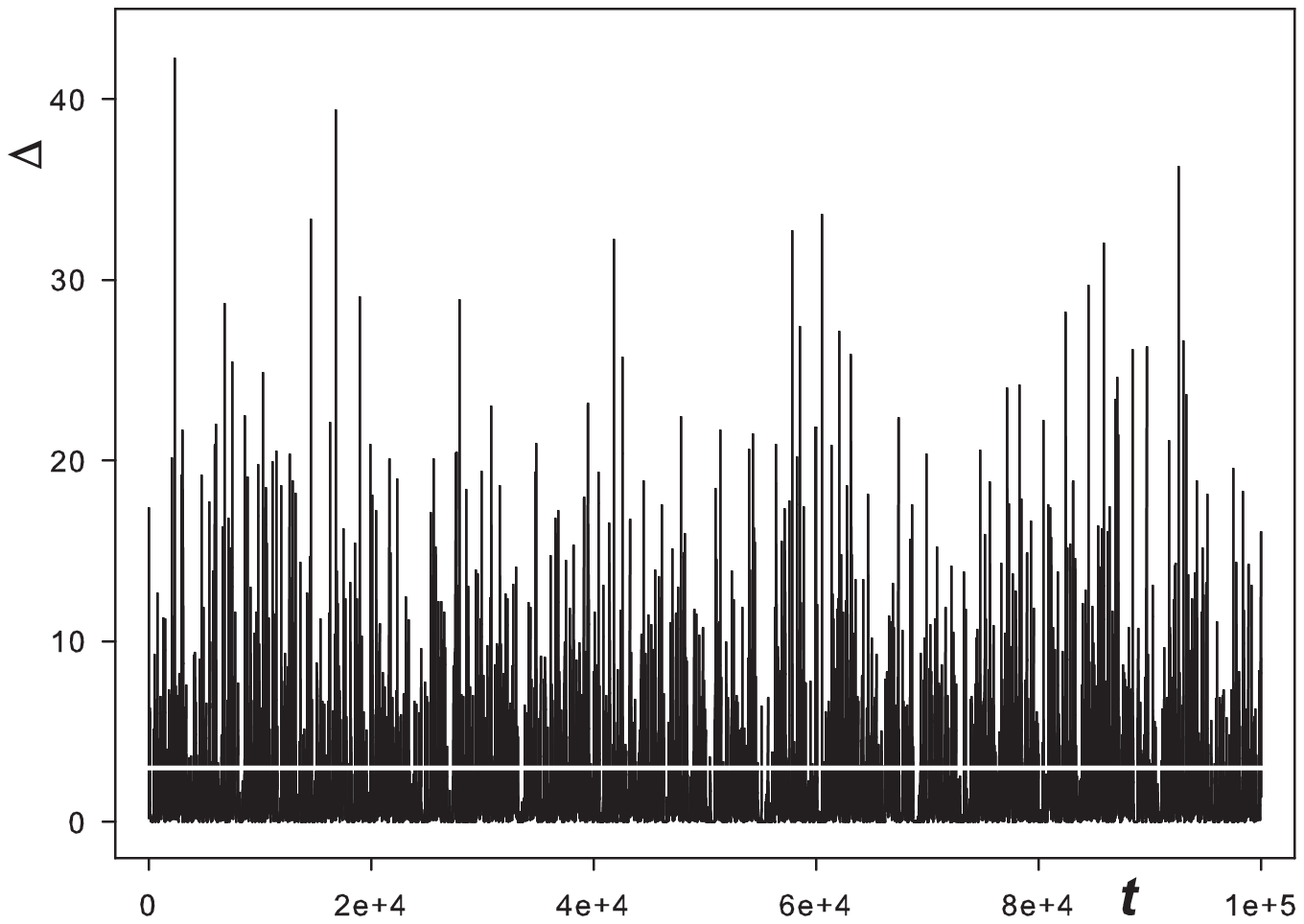}
\end{center}
 \caption{
  \label{fig_10}
Dependence of ${\Delta}(t) =  [\widetilde x_1^1(t)]^2$ on time at
different temperatures of process ${\bf x}^0(t)$. Left panel:
$T=4{.}3$; right panel:  $T = 7{.}0$. $N=5$, integration step $h =
0{.}01$. Time average $\left. \left< \Delta(t) \right>
\right|_{t=0}^{t=10^5}$ are shown in horizontal solid lines.
 }
\end{figure}

Dynamics of process ${\widetilde {\bf x}}^1(t)$ at high
temperatures $T$ was analyzed above in terms of time-average
correlators. And time-resolved behavior of process ${\widetilde
{\bf x}}^1(t)$ at two temperatures $T$ of the background process
${\bf x}^0$ is shown in Fig.~\ref{fig_10}. As above, ${\Delta}(t)
= [\widetilde x_1^1(t)]^2$ was calculated.

One can see that ${\Delta}(t)$ behaves highly irregular. And
numbers and heights of observed peaks increase with the growth of
temperature until the process becomes chaotic at high $T$. At
temperatures $T < T_{\rm thr}$ process $\widetilde {\bf x}_1^1(t)$
consists of individual rare peaks what can point to the
possibility that the energy can be transmitted by impulses.

%%%%%%%%%%%%%%%%%%%%%%%%%%%%%%%%%%%%%%%%%%%%%%%%%%%%%%%%%

\section{Sound, solitons and breathers in the $\beta$-FPU lattice}
  \label{sec:sol_breath}

\begin{figure}
\begin{center}
\includegraphics[width=90mm,angle=0]{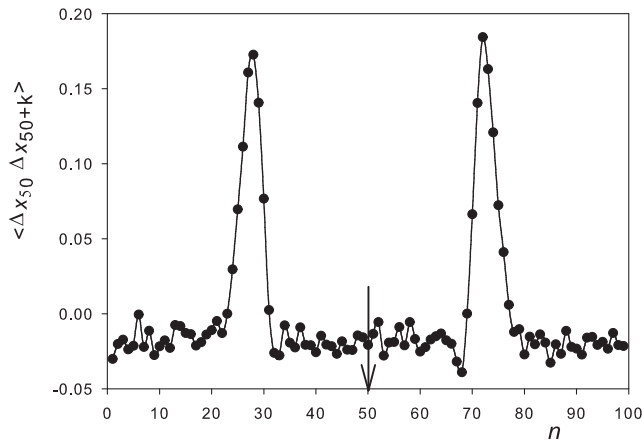}
\end{center}
 \caption{
  \label{fig_11}
Correlator $y_{50}(t) \, y_{50+m}(t+20) $ as a function of the
lattice coordinate $n$. $\beta$-FPU lattice, $N = 101$, $T=2$.
Arrow points to $n=50$
 }
\end{figure}

In accordance with the conjecture on the soliton contribution to
the heat conductivity \cite{Li_05, Vil02}, an attempt was made to
shed some light on this problem. The spatiotemporal correlator
$[y_k(t) \, y_{k+m}(t + \tau)]$ was calculated, where $y_i(t) =
x_i(t) - x_{i-1}(t)$ is the relative displacement of neighboring
particles in time instant $t$. Solitons, being highly correlated
displacements of particles, can leave a trace in correlation
function. Time shift $\tau = 20$ was fixed and spatial correlation
were calculated. Result is shown in Fig.~\ref{fig_11}.
Two peaks in the correlation function, shifted by $m \approx \pm
25$, are visible. The velocity of their propagation is $v_{\rm p}
= m/\tau \approx 1{.}25$.

\subsection{Sound velocity in the $\beta$-FPU lattice}

The temperature dependence of the sound velocity in the
$\beta$-FPU lattice was discovered about a decade ago
\cite{Lep98}. The sound velocity was estimated $v_{\rm snd} \sim
\sqrt{1 + \alpha}$, where $\alpha$ -- parameter of renormalized
frequencies depending on the temperature. Asymptotic value of the
sound velocity in the high temperature limit $v_{\rm snd} \approx
1{.}22 \, T^{1/4}$ was derived recently in \cite{Li_10}. If this
formula apply to $T =2$ then $v_{\rm snd} \approx 1{.}45$ what
differs from $v_{\rm p} \approx 1{.}25$ found from correlation
functions above.

Below we derive more accurate expression for the sound velocity at
low temperatures. It was shown \cite{Ala01} that in nonlinear
systems there exists a spectrum of frequencies which are
proportional to the harmonic ones, according to a well defined
law. Then the $\beta$-FPU potential can be rewritten as
\begin{equation}
  \label{f1}
  u(y) = \left( 1 + \dfrac12 y^2 \right)\dfrac12 y^2
\end{equation}
and an expression in brackets can be replaced by an effective
rigidity $k_{\rm eff}$
\begin{equation}
  \label{f2}
  u(y) = k_{\rm eff}\dfrac{1}{2} y^2.
\end{equation}
As a result the lattice becomes harmonic and it is necessary to
find $k_{\rm eff}$. It can be done in terms of mean field
approximation (MFA). Mean value of potential energy is
\begin{equation}
  \label{f3}
  \left< u_{\rm p}(y) \right> = k_{\rm eff}\dfrac{1}{2} \left< y^2
  \right>,
\end{equation}
where $\left<  y^2 \right>$ is the mean of $y^2$. According to the
virial theorem, mean values of potential and kinetic energies are
equal in the harmonic lattice, that is $\left< u_{\rm p} \right> =
\left< u_{\rm k} \right>$. But the identity $\left< u_{\rm k}
\right> \equiv T/2$ holds for 1D systems. Then
\begin{equation}
  \label{f5}
   k_{\rm eff} \dfrac12 \left< y^2 \right> = \dfrac{T}{2}.
\end{equation}
The condition of self consistency of the MFA is
\begin{equation}
  \label{f6}
  \left(  1 + \dfrac12 \left< y^2 \right>  \right) = k_{\rm
  eff}
  \end{equation}
and it follows that $\left<  y^2 \right> = T/k_{\rm eff}$ and
substitution of this expression into \eqref{f6} gives the
self-consistent equation for $k_{\rm   eff}$:
\begin{equation}
  \label{f7}
  1 + T/(2 k_{\rm eff}) = k_{\rm   eff}
\end{equation}
with the solution
\begin{equation}
  \label{f8}
  k_{\rm   eff} = \dfrac12 + \sqrt{\dfrac14 + \dfrac{T}{2}}.
\end{equation}
Eq.~\eqref{f8} defines the rigidity coefficient for the harmonic
lattice with the renormalized spectrum depending on temperature
$T$. Thereby the temperature renormalized sound velocity
\begin{equation}
  \label{f9}
v_{\rm snd} = \sqrt{k_{\rm eff}} = \sqrt{\frac12 + \sqrt{\frac14 +
\frac{T}{2}}}, \ \ (m=1)
\end{equation}
and $v_{\rm snd} = 1{.}27$ for $T = 2$ what coincides with good
accuracy with $v_{\rm p} \approx 1{.}25$ found from correlation
functions. The high temperature asymptotic of the sound velocity
$v_{\rm snd} \sim 0{.}84 \, \left. T^{1/4} \right|_{T \gg 1}$. The
temperature dependence of sound velocity for temperatures in the
range $0 \le T \le 10$ is shown in Fig.~\ref{fig_13} and very good
agreement between analytical and ``experimental'' results is
observed.
\begin{figure}
\begin{center}
\includegraphics[width=90mm,angle=0]{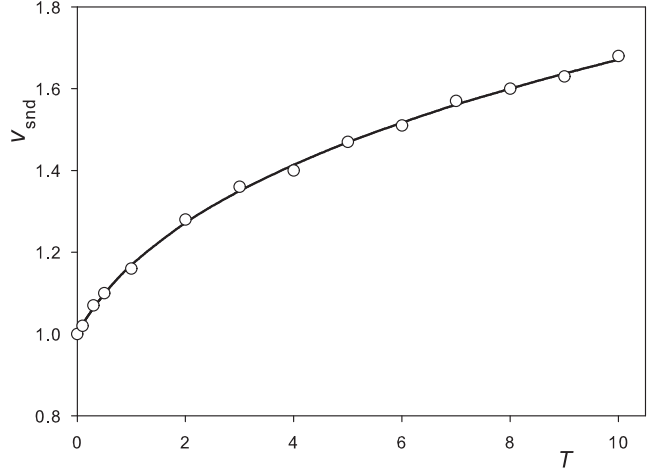}
\end{center}
 \caption{
  \label{fig_12}
The temperature dependence of sound velocity $v_{\rm snd}$: solid
line -- dependence \eqref{f9}; empty circles -- MD simulation.
   }
\end{figure}

If the mean field approximation is applied to the lattices with
cubic nonlinearity $u(y) = \frac12 y^2 \pm \frac13 y^3$ then no
dependence of the sound velocity on temperature is expected.
Really, $u(y) = \frac12 \left( 1 \pm \frac23 y \right) y^2$ and
$k_{\rm eff} = \left( 1 \pm \frac23 \left< y \right> \right) =1$
as $\left< y \right> = 1$.

The obtained results point to the fact that the nonlinearity can
be also ignored in describing the thermodynamics of the
$\beta$-FPU lattice, at least at $T < 10$, and the harmonic
lattice with renormalized rigidity coefficients \eqref{f8} is an
adequate model. This conjecture was checked at different
temperatures $T = 1, 2, 5, 10$ by the comparison of the total
energies computed in MD simulation of the $\beta$-FPU lattice and
its harmonic model with renormalized rigidity coefficients. Very
good coincidence of both energies testifies this hypothesis.

There exists an accurate expression for the specific potential
energy (mean potential energy of one oscillator) \cite{Lik09}
\begin{equation}
  \label{thermodyn}
 \dfrac{\left<U_{\rm p} \right>}{N} = \dfrac18
 \left[ \dfrac{K_{5/4}(q) + K_{3/4}(q)}{2 \, K_{1/2}(q)} -1
 \right]; \qquad q = 1/(8T)
\end{equation}
derived from the thermodynamics of the $\beta$-FPU lattice; $K$ --
modified Bessel functions. Specific potential energy
\eqref{thermodyn} and $\left<U_{\rm p} \right>/N$ computed in
harmonic approximation with rigidity coefficients \eqref{f8}
coincidence with good accuracy.

\subsection{Solitons and breathers in the $\beta$-FPU lattice}

The discrete $\beta$-FPU lattice can be reduced to the modified
Korteweg -- de~Vries (mKdV) equation in the continuum
long-wavelength approximation (see Appendix). The mKdV equation
has solutions in the form of solitons and moving breathers
\cite{Lam80}.

Solitons of compression and elongation has the form
\begin{equation}
  \label{sol}
   y(z,t) = \pm \dfrac{1}{\sqrt{6}} \, B \,
   {\rm sech} \left\{  B \left[ z - \left( 1 + \dfrac{B^2}{24}  \right) t
   \right]  \right\}
\end{equation}
where plus/minus signs stand for elongation/compression solitons;
$B$ -- single parameter which simultaneously determines amplitude,
width and velocity of soliton; $y$ -- local deformation of the
lattice; $z$ -- soliton coordinate at time $t$.

The two-parameteric breather solution is
\begin{equation}
  \label{br}
  y(z,T) = -4\beta\,
\mathrm{sech}{\Psi} \,
\displaystyle\frac{\cos{\Phi}-(\beta/\alpha)\sin{\Phi} \,
\mathrm{tanh} {\Psi}} {1+(\beta/\alpha)^2\sin^2{\Phi} \,
\mathrm{sech} {\Psi}},
\end{equation}
where
\begin{equation}
  \label{phi_psi_2}
  \begin{array}{ll}
  \Psi \equiv 2 \beta (z - \gamma t) + \psi,       \quad &
  \Phi \equiv 2 \alpha (z - \delta t) + \phi , \qquad {\rm and} \\
  \gamma = 4 \, (3 \alpha^2 -\beta^2), &    \delta = 4\, (\alpha^2 -3\beta^2)
\end{array}
  \end{equation}
and $\alpha, \beta$ are free parameters (see Appendix for more
details). It is necessary to make transformation from continuous
variables to discrete variables $y \to x_i - x_{i-1}$ and $z \to
i$ in an attempt to use soliton and breather solutions on the
discrete $\beta$-FPU lattice.

If soliton \eqref{sol} and breather \eqref{br} solutions exist in
the continuum limit, then the question arises: whether these
moving localized excitations can be observed on discrete lattice?
The answer is `yes' and below the visualization method is
suggested.

\begin{figure}
\begin{center}
\includegraphics[width=120mm,angle=0]{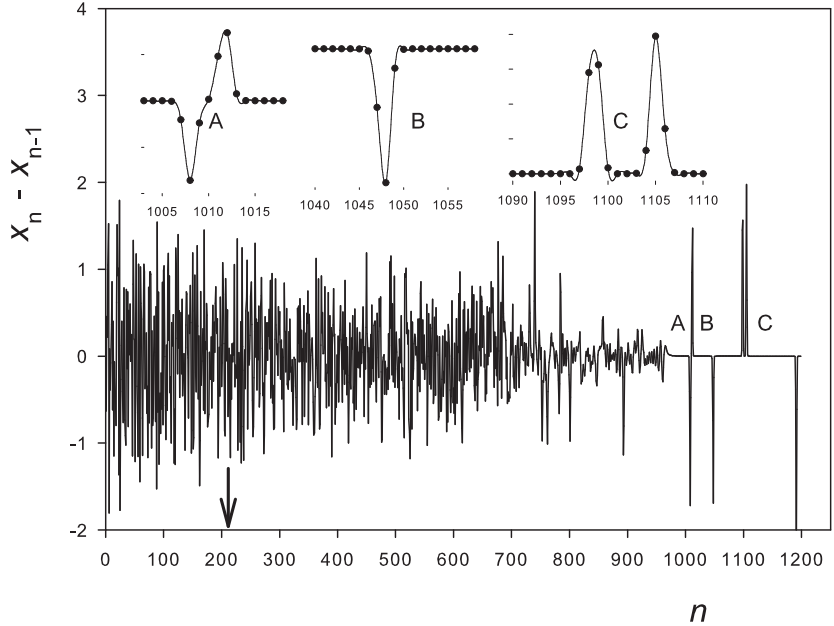}
\end{center}
 \caption{
  \label{fig_13}
Solitons and breather running out of the initially thermalized
lattice. A -- breather, B --  soliton of compression, C -- pair of
antisolitons (solitons of elongation). Arrow at $n = 200$ shows
the border separating initially thermalized and ``cold'' parts of
the lattice. Initial temperature of the left lattice part ($1 \leq
n \leq 200$), $T = 10$.
 }
\end{figure}

Recall that the visualization method for standing discrete
breathers is well known \cite{Bik99, Aub06}: boundaries with
friction forces absorb thermal noise (phonons) and standing
breathers can be easily seen. Other method is necessary to
visualize the moving excitations. Let we have the thermolized
lattice with $N$ oscillators at temperature $T$. If ``cold''
lattice (with zero velocities and displacements) is switched to
the thermalized lattice then solitons and breathers should ``run
out'' to the cold lattice and could be observed. Results are shown
in Fig.~\ref{fig_13} and solitons and breather are immediately
seen.

Solitons and breather (shown in inserts to Fig.~\ref{fig_13}) move
faster then the sound front. At the first glance it is
inconsistent with the relation between velocities of sound and
solitons. The sound velocity at $T = 10$ is $v_{\rm snd} \approx
1{.}67$ what is larger then the maximal soliton velocity $(v_{\rm
sol})_{\rm max} \approx 1{.}3$ (see Appendix). But the temperature
of expanding thermal excitations gradually decreases and the sound
velocity also decreases according to \eqref{f9}. And there comes a
point when solitons, which have constant velocity, keep ahead the
sound front.

There are good grounds for believing that solitons and breathers
do exist in the $\beta$-FPU lattice. Very likely that the soliton
contribution to the heat conductivity increases with the growth of
temperature. Really, the temperature dependence of the soliton
density obeys the relation $n(T) \sim T^{1/3}$ for the Toda
lattice at low temperatures \cite{Mar91}. Conceivably the growth
of the solitons density with temperature might be an inherent
property of nonlinear systems.

Our results on the soliton contribution to the heat conductivity
are inconsistent with previous publication \cite{Li_10} where the
energy carriers are effective phonons rather than solitons.

The possibility of energy transfer by solitons was conjectured
three decades ago \cite{Tod79}. Less studied is the possibility of
energy transfer by breathers. One suggested mechanism is the
Targeted Energy Transfer \cite{Kop01, Man04} when an efficient
energy transfer can occur under a precise condition of nonlinear
resonance between discrete breathers. Various aspects and possible
applications of energy transfer by breathers are considered in
\cite{Aub06}.

%%%%%%%%%%%%%%%%%%%%%%%%%%%%%%%%%%%%%%%%%%%%%%%%%%%%%%%%%

\section{Conclusions}
  \label{sec:concl}

In conclusion we briefly summarize our results. A new method for
the calculation of the heat conductivity is suggested. This is
done by the decomposing of the total dynamics into two parts:
equilibrium process ${\bf x}^0(t)$ at equal temperatures $T$ of
both lattice ends, and nonequilibrium process ${\bf x}^1(t)$ at
temperature $\Delta T$ of one end and zero temperature of the
other. This approach allows to extract and analyze the heat
conductivity in an explicit form.

The primary goal of the paper was to develop a method which would
allow to decrease the computational time at small temperature
gradients when fluctuations of the heat flux are usually too
large. It was supposed that at small temperature gradients, when
the harmonic approximation is valid and an expression for the heat
flux has the form \eqref{2-91}, an analytical averaging over
random Langevin noise terms  can be done. This approach is very
efficient for the calculation of quadratic in ${\bf x}^1(t)$ terms
-- the gain was thousand-fold. But formulaes are very complex for
the linear terms and an efficient algorithm for their realization
was not found yet. By this expedient the objective has not been
met in full: the gain in computational time is obvious for small
$(N \lesssim 100)$ lattices, but decreases as the lattice length
increases. Nevertheless we suppose that the further analysis of
process ${\bf x}^1(t)$ can be useful as it is responsible for the
energy transfer.

The threshold phenomena are familiar in microcanonical ensembles
\cite{Lic08}. There exists two values of specific energy $E$. One
separates dynamical regime and weak chaos, and higher $E$
separates weak and strong chaos. It may be inferred that an
existence of threshold phenomena is also a common occurrence in
canonical ensembles. Really, a threshold temperature $T_{\rm thr}$
was found. The threshold temperature separates the different
behavior of process ${\bf x}^1(t)$: process ${\bf x}^1(t)$ damps
out at $T < T_{\rm thr}$ and reaches the stationary value at $T >
T_{\rm thr}$. It may be conceived that the soliton and breather
contributions to the heat conductivity increases with the growth
of temperature if $T > T_{\rm thr}$. Solitons and breathers can
emerge from either thermal fluctuations or higher order phonon
interactions. Additional experiments for nanosized systems at low
temperatures when $T < T_{\rm thr}$ can reveal some new features
omitted in the present work.

The modified Korteweg -- de~Vries equation is derived in the
continuum approximation for the $\beta$-FPU lattice. mKdV has
solutions in the form of compression/elongation solitons and
breathers. The stability of these quasi-particles was checked in
numerical experiments. Both types of excitations were directly
visualized.

On the other hand, it was found that the non-linear $\beta$-FPU
lattice can be reduced to the harmonic lattice with the
temperature renormalized frequency spectrum. This reduction allows
to reproduce adequately the heat conductivity and thermodynamics
of the parent lattice. These two, mutually contradictory,
properties of the $\beta$-FPU lattice, -- an inherent existence of
solitons and breathers, and its ``harmonic'' behavior, seem to be
very strange. Further analysis is necessary to solve this dilemma.

It is likely that some fraction of total heat conductivity is
conditioned by solitons and moving breathers. But their
contribution to the energy transfer deserves further
investigation.

The $\beta$-FPU lattice is unique in the sense that in the
continuum limit it has stable solutions in the form of solitons of
compression and elongation. If the number of compression and
elongation solitons is equal on average, then no macroscopic
changes in the lattice length appear and an additional energy of
deformation is negligible. It is an additional energetic factor
favoring the solitons existence.

The case is quite different for unsymmetrical potentials which in
the lowest order of the Taylor expansions have the form $u(y) =
\frac{a}{2}y^2 \pm \frac{b}{3}y^3$. $\alpha$-FPU, Toda, Morse,
Lennard-Jones potentials are the examples. All these ``cubic''
potentials can be easily reduced to the ordinary KdV equation with
the solution in the form of soliton of compression. And the large
number of solitons is highly energetically unfavorable due to
macroscopical compression of the lattice length.

%\begin{acknowlegments}
%The authors thank the anonymous reviewers for few valuable
%comments.
%\end{acknowlegments}

%%%%%%%%%%%%%%%%%%%%%%%%%%%%%%%%%%%%%%%%%%%%%%%%%%%%%%%%%
\appendix

\section{The modifief Korteweg -- de~Vries equation \\
for the $\beta$-FPU lattice. Solitons and breathers.}

An approximate solution for the soliton of compression in the
$\beta$-FPU lattice was obtained about two decades ago
\cite{Wat93}. Analogous soliton solution with the profile $Q_{\rm
snd} = \sqrt{2 \, (v_{\rm snd}^2 -1)} \, {\rm sech} \left[2 \, z
\sqrt{(v_{\rm snd}^2 -1)/v_{\rm snd}^2} \, \right]; \quad (z \sim
x_i - x_{i-1})$ was used recently \cite{Li_10} and this solution
is the function of a single parameter -- sound velocity $v_{\rm
snd}$.

Below we derive an equation for the continuum analogue of the
$\beta$-FPU lattice with more accurate soliton and breather
solutions. The $\beta$-FPU potential has general form
\begin{equation}
  \label{A1}
  u(y_i) = \dfrac{\alpha}{2} \, y_i^2 + \dfrac{\beta}{4} \, y_i^4;
  \quad (m = 1),
\end{equation}
where $y_i = x_i - x_{i-1}$ is the relative displacement of
neighboring oscillators. The corresponding equations of motion are
\begin{equation}
  \label{A2}
  \ddot y_i = \alpha \, \left( y_{i-1} - 2 y_i + y_{i+1} \right) +
              \beta \,\left( y_{i-1}^3 - 2 y_i^3 + y_{i+1}^3 \right).
\end{equation}

Starting from \eqref{A2}, the continuum approximation can be
derived supposing small deviations from equilibrium. The series
expansion in terms of $y_i$ up to the forth order is:
\begin{equation}
  \label{A3}
  y_{i \pm 1} = y_i \pm y_i' + \dfrac12 y_i'' \pm \dfrac16 y_i'''
  + \dfrac{1}{24} y_i^{IV} \,.
\end{equation}
Substitution of this expansion into \eqref{A2} gives the continuum
equation:
\begin{equation}
  \label{A4}
 \ddot y = \alpha \left( y'' + \dfrac{1}{12} y^{IV} \right) +
 3 \beta \left( y^2 \, y' \right)^2 \,.
\end{equation}
Below we follow the well known reductive perturbation method (RPM)
\cite{Sas81, Leb08} to get necessary equation in partial
derivatives. The receipt consists in introducing new variables
\begin{equation}
  \label{A5}
  \begin{split}
   y    & = \varepsilon^{1/2} u_1 + \varepsilon^{3/2} u_2 + \ldots \,; \\
   \xi  & = \varepsilon^{1/2} (z - ct); \\
   \tau & = \varepsilon^{3/2} t \,.
  \end{split}
\end{equation}
Next the hierarchy of the expansions in terms of small parameter
$\varepsilon$ should be used. If \eqref{A5}  is substituted in
\eqref{A4} and terms of the order $\varepsilon^{3/2}$ and higher
are neglected, then $ c^2(u_1)_{\xi \xi} = \alpha (u_1)_{\xi \xi}$
and $c = \sqrt{\alpha}$. $c$ is the sound velocity in the harmonic
approximation. Transformation \eqref{A5} means that the new
coordinate system $\xi$ moves with velocity $c$ relative to the
old coordinate system $z$.

Equation with the accuracy of the order $\varepsilon^{5/2}$ is
\begin{equation}
  \label{A6}
  2 c (u_1)_{\tau} + 3 \beta (u_1)^2(u_1)_{\xi} + \dfrac{\alpha}{12}
  (u_1)_{\xi \xi \xi} =0
\end{equation}
and one can see that \eqref{A6} reminds the well known modified
KdV equation. Additional variables substitution $u_1 =
\sqrt{\alpha/6\beta} w$ and $\tau= (24/\sqrt{\alpha}) T$ should be
done to get the exact form of the mKdV equation:
\begin{equation}
  \label{A7}
  w_T + 6  w^2 w_{\xi} + w_{\xi \xi \xi} =0 \,.
\end{equation}
$\xi$ and $T$ are spatial and time variables.

Equation \eqref{A7} has two types of solutions \cite{Lam80}. The
soliton solution is
\begin{equation}
  \label{A8}
  w(\xi, T) = \pm B \, {\rm sech} (B \xi - B^3 T + \delta)  =
  \pm B \, {\rm sech} [B (\xi - B^2 T) - \xi_0] \,.
\end{equation}
Plus/minus signs are related to elongation/contraction solitons,
correspondingly. Soliton  \eqref{A8} is the one-parametric
solution: free parameter $B$ defines simultaneously amplitude
($B$), half-width ($\sim 1/B$) and velocity ($B^2$); $\xi_0$
defines the soliton coordinate at $T = 0$.

Returning back to initial coordinates $\left\{ z,t \right\}$, the
soliton solution is
\begin{equation}
  \label{A11}
  y(z,t) = \pm  \sqrt{\dfrac{\alpha}{6 \beta}} \, B \,
  {\rm sech} \left\{ B \left[ z - \left( 1 + \sqrt{\alpha}{B^2 }/{24}
  \right)  t \right] - z_0 \right\},
\end{equation}
where $z$ -- coordinate, $t$ -- time, $z_0$ -- soliton coordinate
at $t=0$. The soliton velocity is $v_{\rm sol} = \sqrt{\alpha}(1 +
B^2/24)$ and is ``supersonic'' relative to the sound velocity in
the harmonic approximation $v_{\rm snd}^0 = \sqrt{\alpha} $. If
$B$ increases then soliton has larger amplitude, becomes more
narrow and its velocity increases. The soliton solution for the
$\beta$-FPU lattice can be written if discrete variables in
\eqref{A11} are used: $y_i = x_i - x_{i-1}$, $z \to i$, $z_0 \to
i_0$.

Parameter $B$ can be arbitrary large in the continuum limit
\eqref{A11}. But the lattice discreteness imposes limitations on
the soliton width: solitons with the half-width less then
$\lesssim 2.0$ become unstable. That is to say, soliton amplitude
and velocity also have upper limit: $A = \sqrt{\alpha/6 \beta}B
\lesssim 1$, \,\, $1 \leq v_{\rm sol} \lesssim 1{.}3$ and the free
parameter $B$ for discrete $\beta$-FPU lattice $B \lesssim \sqrt{6
\beta/\alpha}$.

Two-parametric breather solution is
\begin{equation}
  \label{A9}
  w(\xi, T) = - 4 \beta \, {\rm sech} \, \Psi
  \left[ \dfrac{\cos \Phi - (\beta/\alpha) \sin \!\Psi \tanh \! \Psi}
  {1 + (\beta/\alpha)^2 \sin^2 \! \Phi \,\,{\rm sech} \Psi}
  \right]\,,
\end{equation}
where
\begin{equation}
  \label{phi_psi}
    \begin{split}
      \Psi \equiv 2 \beta \xi  - \gamma T - \psi, \quad &
      \Phi \equiv 2 \alpha \xi - \delta T - \phi; \\
      \gamma = 8 \beta (2 \alpha^2 - \beta^2), \quad &
      \delta = 8 \alpha (\alpha^2 - 3 \beta^2)
    \end{split}
\end{equation}
$\alpha, \, \beta$ are free parameters; $\psi, \, \phi$ -- initial
phases; the group and phase velocities are $v_{\rm gr} =
\gamma/2\beta$ and $v_{\rm ph} = \delta/ 2 \alpha $,
correspondingly. Returning back to coordinate system $z$, the
breather solution takes the form
\begin{equation}
  \label{breath}
  y(z,t) = -\sqrt{\dfrac{8 \alpha}{3 \beta}} \, \beta \,\,
           {\rm sech} {\widetilde \Psi} \,
           \dfrac{ \cos {\widetilde \Phi} - (\beta/\alpha)
           \sin \! {\widetilde \Phi} \,\,{\rm tanh} {\widetilde \Psi} }
           { 1 + (\beta/\alpha)^2 \sin^2 \!{\widetilde \Phi} \,\,
           {\rm sech}{\widetilde \Psi} } \,,
\end{equation}
where
\begin{equation}
  \label{par}
   {\widetilde \Psi} \equiv 2 \beta
   \left[z - \sqrt{\alpha}\left( 1 - \dfrac{\gamma}{48 \beta}  \right)
    t\right], \quad
   {\widetilde \Phi} \equiv 2 \alpha
   \left[z - \sqrt{\alpha}\left(1 - \dfrac{\delta}{48 \alpha} \right)
     t \right]
\end{equation}
and $\gamma, \delta$ are defined in \eqref{phi_psi}.

\begin{figure}
  \begin{center}
  \includegraphics[width=75mm,angle=0]{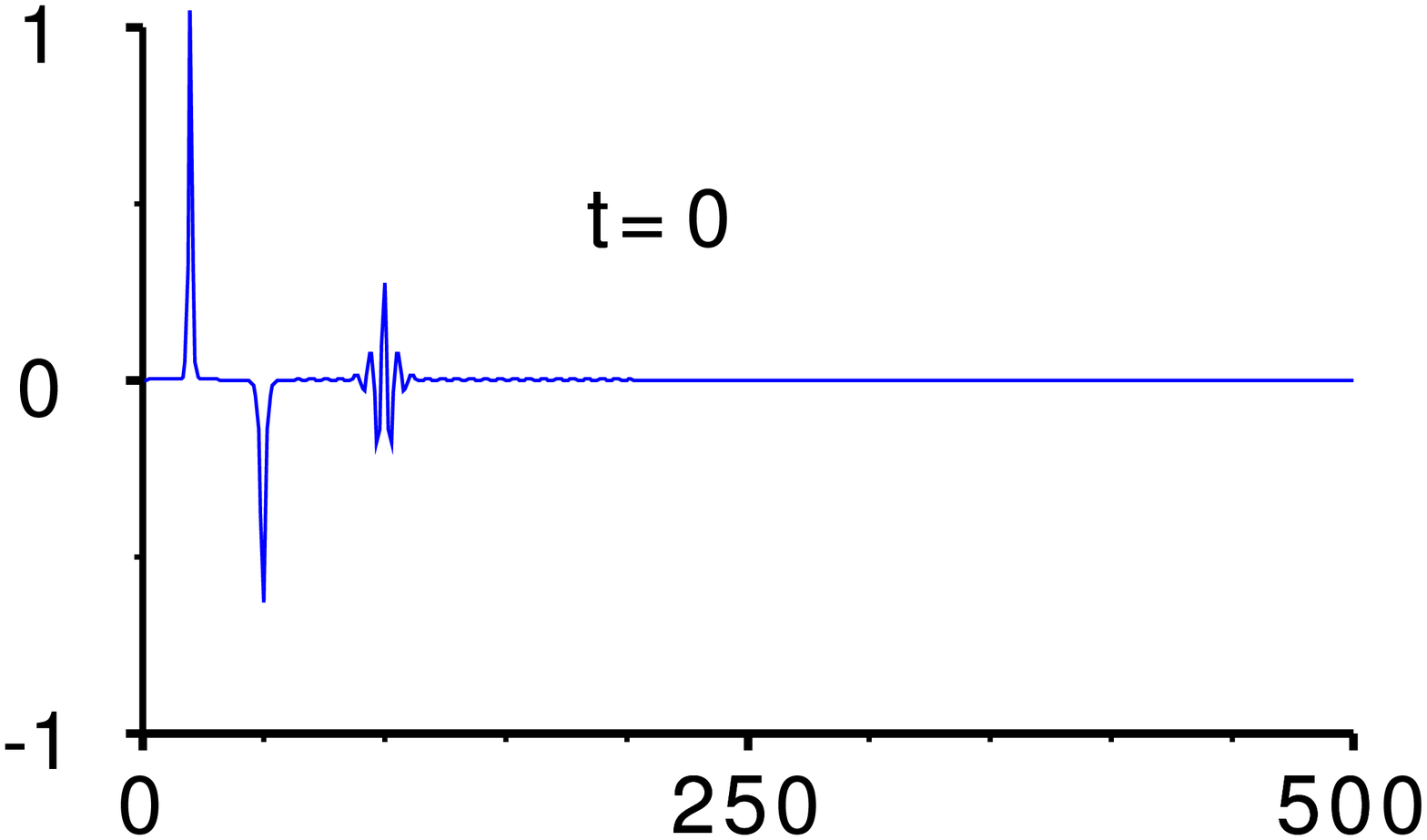}
  \includegraphics[width=75mm,angle=0]{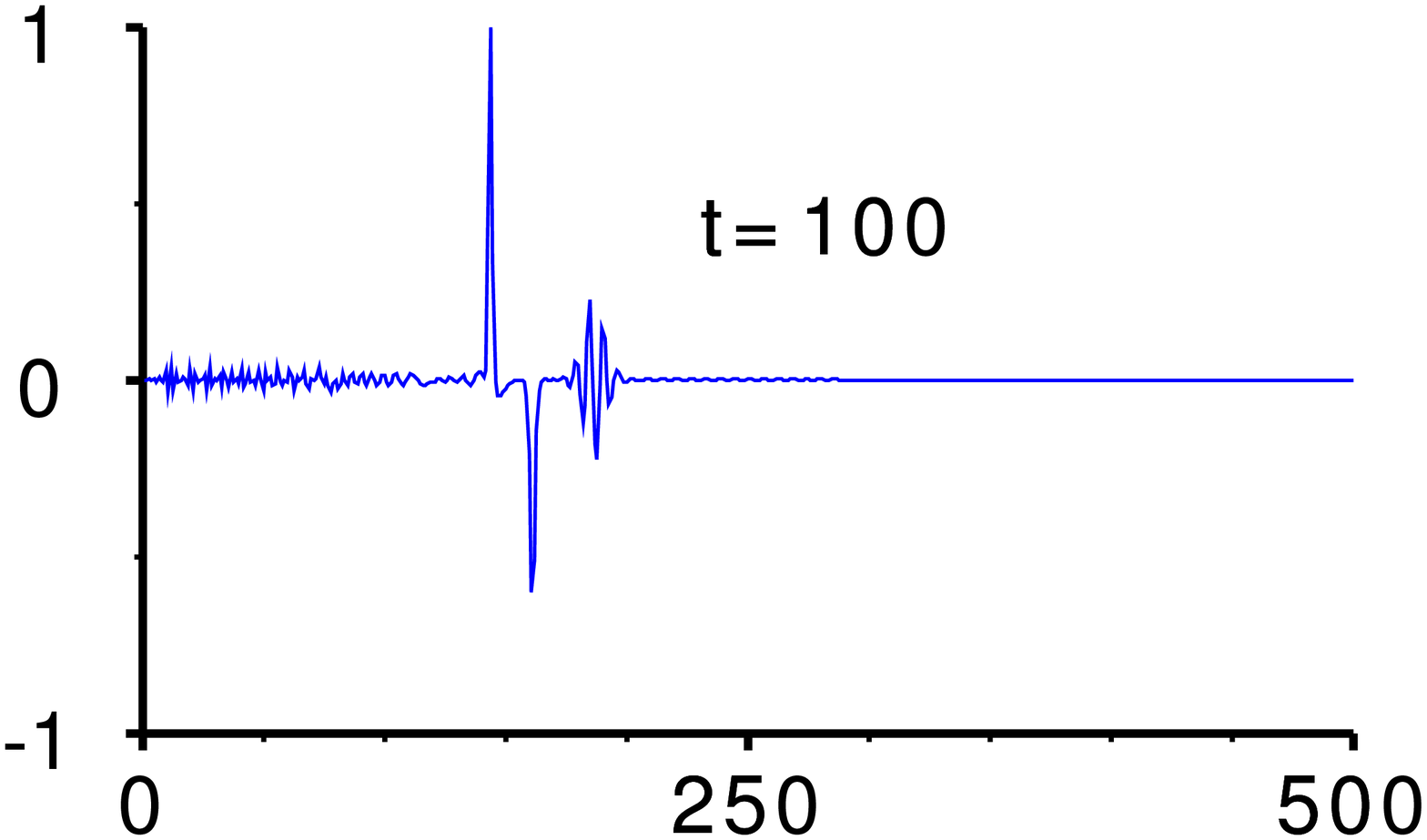}
\hspace{8 cm} {\large a} \hspace{7 cm} {\large b}

\includegraphics[width=75mm,angle=0]{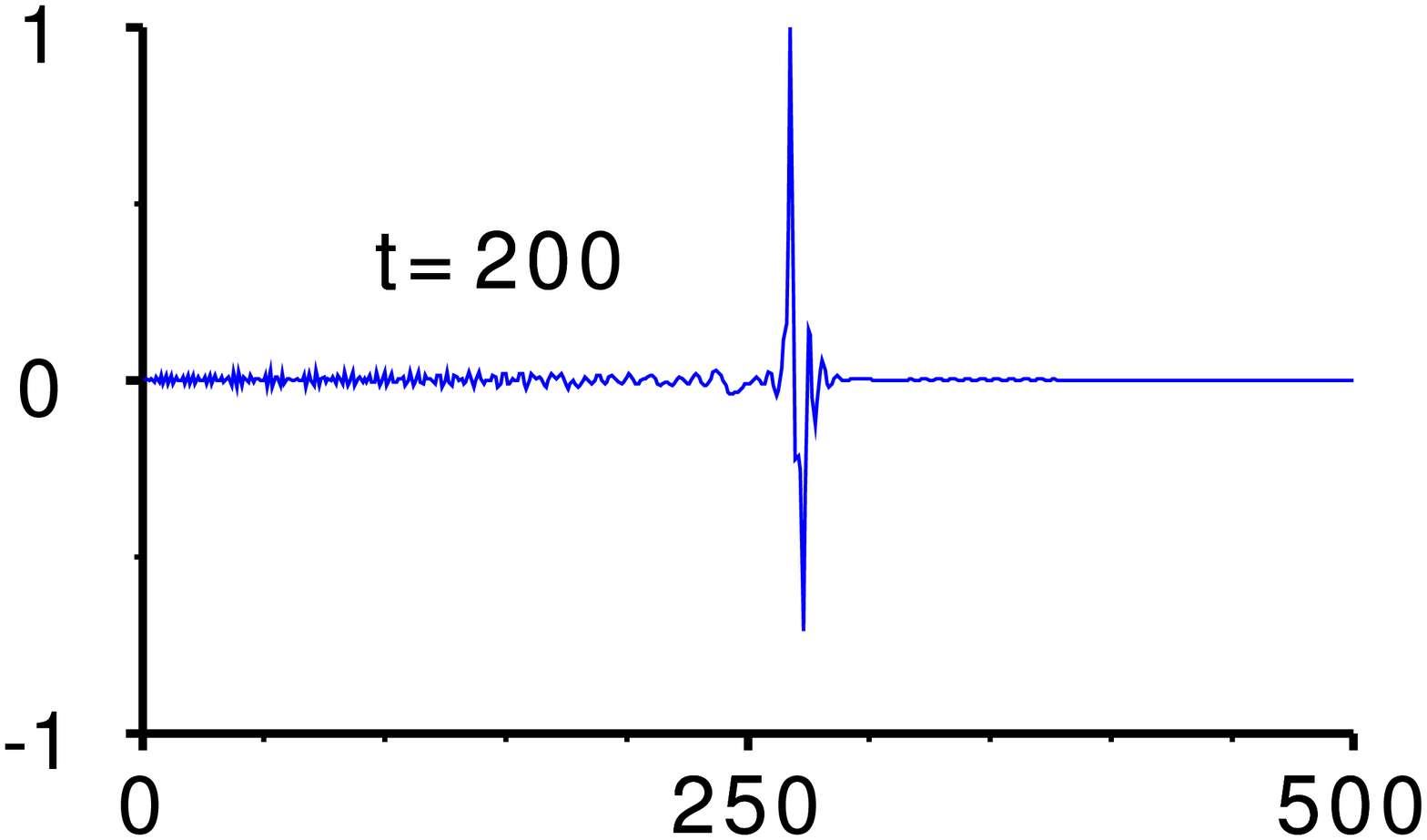}
  \includegraphics[width=75mm,angle=0]{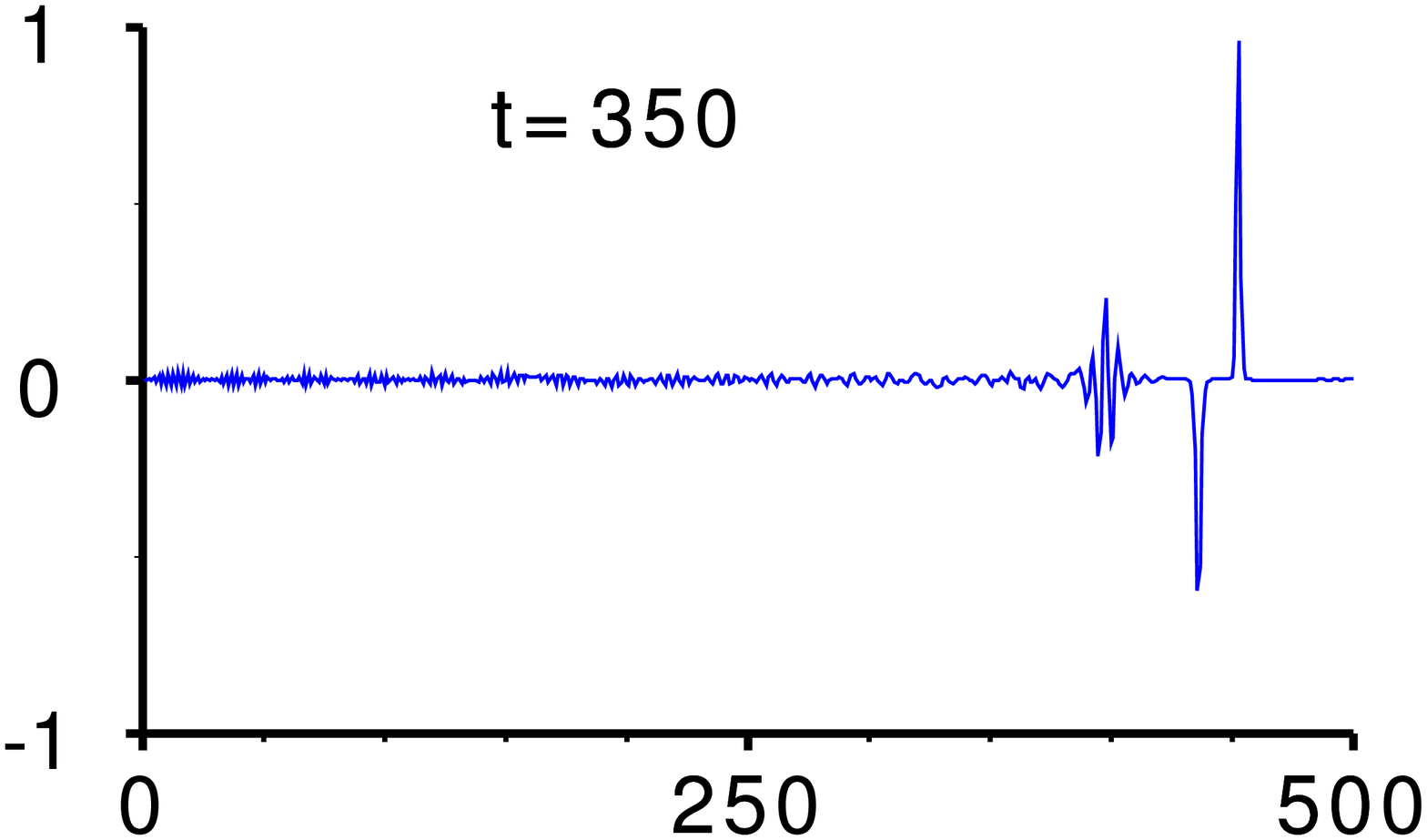}
\hspace{8 cm} {\large c} \hspace{7 cm} {\large d}

  \end{center}
  \caption{  \label{fig_14}
    Movement and collision of two solitons and one breather.
    Snapshots are made at $t = 0,\, 100,\, 200,\,  350$ t.u.
         }
\end{figure}

The group breather velocity $v_{\rm gr} = \sqrt{\alpha}(1 -
\gamma/48 \beta)  = \sqrt{\alpha}\,[1 - (2 \alpha^2 -
\beta^2)/6]$. Its amplitude $\sim \sqrt{8 \alpha \beta/ 3}$.

Soliton and breather stability was checked in numerical
simulation. Initial conditions are chosen in the form of two
solitons and one breather. Relative displacements are chosen
according to \eqref{A11} and \eqref{A9}, correspondingly.
Velocities were obtained by differentiating by time.

The most left soliton of elongation has amplitude $A_1 \simeq
1.09$ and velocity $v_1 \simeq 1.3$ and initially located at $i_1
= 20$. Next soliton of compression has amplitude $A_2 \simeq 0.63$
and velocity $v_2 \simeq 1.1$ and initially located at $i_2 = 50$.
Breather with parameters $\alpha = \pi/6, \,\, \beta = 1/6$ is
initially centered at $i_3 = 100$ and its velocity $v_3 = 0.91$
(see Fig.~14a). The velocities condition $v_1 > v_2 > v_3$ says
that solitons and breather should meet and interact. This triple
interaction is shown in Fig.~14c. But as result, all three species
preserve their individuality (Fig.~14d). These findings (large
free paths, preserving amplitudes and shapes after collision)
unambiguously demonstrate that soliton and breather are do really
exist in the $\beta$-FPU lattice

%%%%%%%%%%%%%%%%%%%%%%%%%%%%%%%%%%%%%%%%%%%%%%%%


\begin{thebibliography}{99}

\bibitem{Lep03} S. Lepri, R. Livi, A. Politi, Thermal conduction in
classical low-dimensional lattices, Phys. Reports 377 (2003)
1--80.

\bibitem{Fer55} E. Fermi, J. Pasta, S. Ulam, Document
LA--1940 (May 1955); Collected papers of E.~Fermi, University of
Chicago Press, Chicago, 1965, vol. 2, p. 78.

\bibitem{Cha05} CHAOS {\bf 15}(1) (2005) (Focus Issue: The Fermi-Pasta-Ulam
Problem -- The First 50 Years, Ed. by D.K. Campbell, Ph. Rosenau,
and G.M. Zaslavsky).

\bibitem{Lec08} G. Gallavotti (Ed.),  The Fermi-Pasta-Ulam Problem: A Status
Report, Lect. Notes Phys. 728, Springer, Berlin Heidelberg, 2008.

\bibitem{Kru64} M.D. Kruskal, N.J. Zabusky,
Stroboscopis-perturbation procedure for treating a class of
nonlinear wave equations, J. Math. Phys. 5 (1964) 231--244.

\bibitem{Zab65} N.J. Zabusky, M.D. Kruskal, Interaction of ``solitons''
in a collisionless plasma and the recurrence of initial states,
Phys. Rev. Lett. 15 (1965) 240--243.

\bibitem{Dod82} R.K. Dodd, J.C. Eilbeck, J.D. Gibbon, H.C. Morris,  Solitons
and Nonlinear Wave Equations, Academic Press, New York, 1982.

\bibitem{Cam96} D.K. Campbell, M. Peyrard,
Heat conduction in a one-dimensional aperiodic system, Physica D
18 (1986) 47--53.

\bibitem{Sie88} A. J. Sievers, S. Takeno, Intrinsic localized modes in
anharmonic crystals, Phys. Rev. Lett. 61 (1988), 970--973.

\bibitem{Dau93} T. Dauxois, M. Peyrard, Energy localization in
nonlinear lattices, Phys. Rev. Lett. 70 (1993) 3935--3938.

\bibitem{Aub94} S. Aubry,
The concept of anti-integrability applied to dynamical systems and
to structural and electronic models in condensed matter physics,
Physica D 71 (1994) 196--221.

\bibitem{Mac94} R.S. MacKay, S. Aubry,
Proof of existence of breathers for time-reversible or hamiltonian
networks of weakly coupled oscillators, Nonlinearity 7 (1994)
1623--1643.

\bibitem{Por09} M.A. Porter, N.J. Zabusky, B. Hu, D.K. Campbell,
Fermi, Pasta, Ulam and the birth of experimental mathematics. A
numerical experiment that Enrico Fermi, John Pasta, and Stanislaw
Ulam reported 54 years ago continues to inspire discovery,
American Scientist 97(3) (2009) 214--221.

\bibitem{Deb14} P.~Debye,  Vortr\"age \"uber die Kinetische Theorie der
W\"arme, Teubner, 1914.

\bibitem{Cas05} G. Casati, B. Li,
Heat conduction in one dimensional systems: Fourier law, chaos,
and heat control, arXiv:cond-mat/0502546.

\bibitem{Lic08} A.J. Lichtenberg, R. Livi, M. Pettini, S. Ruffo,
Dynamics of oscillator chains,  Lect. Notes Phys. 728 (2008)
21--121 .

\bibitem{Bar07}. D. Barik, Anomalous heat conduction in
a 2d Frenkel-Kontorova lattice, Europ. Phys. J. B 56 (2007)
229--234.

\bibitem{Lip00} A. Lippi, R. Livi, Heat-conduction in
two-dimensional nonlinear lattices, J. Stat. Phys. 100 (2000)
1147--1172.

\bibitem{Sai10} K. Saito, A. Dhar, Heat conduction in a
three dimensional anharmonic crystal, Phys. Rev. Lett. 104 (2010),
040601.

\bibitem{Shi08} H. Shiba, N. Ito,
Anomalous heat conduction in three-dimensional nonlinear lattices
J. Phys. Soc. Jap. 77 (2008) 054006.

\bibitem{Hen09} A. Henry, G. Chen, Anomalous heat conduction in polyethylene
chains: Theory and molecular dynamics simulations, Phys. Rev. B
 79 (2009) 144305/1--10.

\bibitem{Yao05} Z. Yao,  J.-S. Wang, B. Li, G.-R. Liu,
Thermal conduction of carbon nanotubes using molecular dynamics,
Phys. Rev B  71 (2005) 085417/1--8.

\bibitem{Yu_05} C. Yu, L. Shi, Z. Yao, D. Li, A. Majumdar,
Thermal conductance and thermopower of an individual single-wall
carbon nanotube, Nano Lett. 5 (2005) 1842--1846.

\bibitem{Mar02} S. Maruyama, A molecular dynamics simulation of
heat conduction in finite length SWNTs, Physica B 323 (2002)
193--195.

\bibitem{Min05} N. Mingo, and D.A. Broido,
Length dependence of carbon nanotube thermal conductivity and the
``problem of long waves'', Nano Lett. 5 (2005) 1221--1225.

\bibitem{Cao04} J.X. Cao, X.H. Yan, Y. Xiao, J.W. Ding,
Thermal conductivity of zigzag single-walled carbon nanotubes:
Role of the umklapp process, Phys. Rev. B  69 (2004) 073407/1--4.

\bibitem{Lep05} S. Lepri, R. Livi, A. Politi,
Studies of thermal conductivity in Fermi-Pasta-Ulam-like lattices,
CHAOS 15 (2005) 015118/1--9.

\bibitem{Tod79} M. Toda,
Solitons and heat conduction, Phys. Scr. 20 (1979) 424--430.

\bibitem{Fri99} G. Friesecke, R.L. Pego, Solitary waves on the
FPU lattices. I. Qualitative properties, renormalization and
continuum limit, Nonlinearity 12 (1999), 1601--1628.

\bibitem{Fri02} G. Friesecke, R.L. Pego, Solitary waves on the
FPU lattices. II. Linear implies nonlinear stability, Nonlinearity
15 (2002), 1343--1360.

\bibitem{Fri04a} G. Friesecke, R.L. Pego, Solitary waves on the
FPU lattices. III. Howland-type Floquet theory, Nonlinearity 17
(2004), 207--228.

\bibitem{Fri04b} G. Friesecke, R.L. Pego, Solitary waves on the
FPU lattices. IV. Proof of stability at low energy, Nonlinearity
17 (2004), 229--252.

\bibitem{Hof08} A. Hoffman, C.E. Wayne,
Asymptotic two-soliton solutions in the Fermi-Pasta-Ulam Model, J.
Dynamics and Diff. Equations 21 (2009) 343--351.

\bibitem{Li_05} B. Li, J. Wang, L. Wang G. Zhang,
Anomalous heat conduction and anomalous diffusion in nonlinear
lattices, single walled nanotubes, and billiard gas channels,
CHAOS  15 (2005) 015121/1--13.

\bibitem{Vil02} H.J. Viljoen, L.L. Lauderback D. Sornette,
Solitary waves and supersonic reaction front in metastable solids,
Phys. Rev. E 65 (2002) 026609/1--13.

\bibitem{Mar03} S. Maruyama,
A molecular dynamics simulation of heat conduction of a finite
length single-walled carbon nanotube, Microscale Thermophysical
Engineering 7 (2003) 41--50.

\bibitem{Aok01} K. Aoki, D. Kusnezov,
Fermi-Pasta-Ulam $\beta$-model: boundary jumps, Fourier's law, and
scaling,  Phys. Rev. Lett. 86 (2001) 4029--4032.

\bibitem{Aok00} K. Aoki, D. Kusnezov,
Bulk properties of anharmonic chains in strong thermal gradients:
non-equilibrium phi(4) theory, Phys. Lett. A 265 (2000) 250--256.

\bibitem{Nar02} O.Narayan, S. Ramaswamy, Anomalous heat
conduction in one-dimensional momentum-conserving systems, Phys.
Rev. Lett. 89 (2002) 200601.

\bibitem{Wan07} J.-S. Wang,
Quantum thermal transport from classical molecular dynamics, Phys.
Rev. Lett. 99 (2007) 160601/1-4.

\bibitem{Zha02} Y. Zhang, H. Zhao,
Heat conduction in a one-dimensional aperiodic system, Phys. Rev.
E  66 (2002) 026106/1-4.

\bibitem{Joh08} M. Johansson, G. Kopidakis, S. Lepri,
S. Aubry, Transmission thresholds in time-periodically driven
nonlinear disordered systems, Europh. Lett. 86 (2009) 10009.

\bibitem{Rie67} Z. Rieder, J.L. Lebowitz, E.Lieb,
Properties of a harmonic crystal in a stationary nonequilibrium
state, J. Math. Phys. 8 (1967) 1073--1078.

\bibitem{Lep98} S. Lepri, Relaxation of classical many-body Hamiltonians in one
dimension, Phys. Rev. E 58 (1998) 7165--7171.

\bibitem{Li_10} N. Li, B. Li, S. Flach,
Energy carriers in the Fermi-Pasta-Ulam $\beta$-lattice: solitons
or phonons?, Phys. Rev. Lett. 105 (2010) 054102/1-4.

\bibitem{Ala01} C. Alabiso, M. Casartelli,
Normal modes on average for purely stochastic systems, J. Phys. A
34 (2001) 1223--1230.

\bibitem{Lik09} V.N. Likhachev, T.Yu. Astakhova, W. Ebeling, G.A. Vinogradov,
Equilibrium thermodynamics and thermodynamic processes in
nonlinear systems, Eur. Phys. J. B 72 (2009) 247--256.

\bibitem{Lam80} G. Lamb, Elements of soliton theory, John Willey \& Sons, New
York, 1980.

\bibitem{Bik99} A. Bikaki, N.K. Voulgarakis, S. Aubry,  G.P. Tsironis, Energy
relaxation in discrete nonlinear lattices, Phys. Rev. E 59 (1999)
1234--1237.

\bibitem{Aub06} S. Aubry,
Discrete Breathers: Localization and transfer of energy in
discrete Hamiltonian nonlinear systems, Physica D 216 (2006)
1--30.

\bibitem{Mar91} F. Marchesoni, C. Lucheroni,
Heat conduction in a one-dimensional aperiodic system, Phys. Rev.
B 44 (1991) 5303--5305.

\bibitem{Kop01} G. Kopidakis, S. Aubry, G.P. Tsironis,
Targeted energy transfer through discrete breathers in nonlinear
systems, Phys. Rev. Lett 87 (2001) 165501/1-4.

\bibitem{Man04} P. Maniadis, G. Kopidakis, S. Aubry,
The concept of anti-integrability applied to dynamical systems and
to structural and electronic models in condensed matter physics,
Physica D 188 (2004) 153--177.



\bibitem{Wat93} J.A. Wattis, Approximations to solitary waves
on lattices. 2. Quasi-continuum methods for fast and slow waves,
J. Phys. A 26 (1993) 1193--1209.

\bibitem{Sas81} K. Sasaki, Solitons in one-dimensional
helimagnets, Progr. Theor. Phys. 65 (1981) 1787--1797.

\bibitem{Leb08} H. Leblond,
The reductive perturbation method and some of its applications, J.
Phys. B 41 (2008) 043001.

\end{thebibliography}
\end{document}